%% file: main.tex
\documentclass[aps,prl,reprint,floatfix,superscriptaddress]{revtex4-2}

\input{header}

\begin{document}

\title{Causality from the spectrum: \\ Emergence of causal order from process-matrix mereology}

\date{\today}

\author{Varun Kushwaha}
\affiliation{Faculty of Physics, Ludwig-Maximilians-Universität, Scheinerstr.\ 1, 81677 Munich, Germany}
\author{Nicolas Loizeau}
\affiliation{Niels Bohr Institute, University of Copenhagen, Copenhagen, Denmark}
\author{Alexei Grinbaum}
\affiliation{CEA-Saclay, 91191 Gif-sur-Yvette Cedex, France}
\author{Oliver Friedrich}
\affiliation{Faculty of Physics, Ludwig-Maximilians-Universität, Scheinerstr.\ 1, 81677 Munich, Germany}
\affiliation{Excellence Cluster ORIGINS, Boltzmannstr.\ 2, 85748 Garching, Germany}

\begin{abstract}
In Hamiltonian systems, the only basis-independent quantity is the spectrum. Given a spectrum, quantum mereology seeks 
preferred tensor-product decompositions into local subsystems,
leading to the emergence of locality. Causal order, however, remains fixed by the Hamiltonian time evolution. 
In contrast, higher-order quantum theory
permits more general processes that need not possess a definite global causal order.
In the process-matrix framework, specifying a process requires a choice of subsystems corresponding to the input and output Hilbert spaces of each agent. A change of basis redefines both the agents and the corresponding decompositions into subsystems, while leaving the spectrum of the process matrix invariant. Here, we study how causality arises from this spectrum. 
First, we derive spectral constraints
on processes compatible with definite causal order.
Second, we show that, in the thermodynamic limit, generic quantum processes admit a preferred decomposition with a definite causal order. 
This suggests a mechanism for the emergence of classical causality only from algebraic 
ingredients. 
\end{abstract}

\maketitle

\section{Introduction}

\subsection{From locality to causality in quantum mereology}
A foundational question in physics is whether spacetime concepts like locality and causality can be derived rather than postulated. From a quantum-mechanical perspective, this translates into 
the question about the origin of the decomposition into subsystems.
Can a ``natural'' or ``preferred'' tensor product decomposition of the global Hilbert space arise only from basis-independent spectral data? Quantum mereology pursues this program \cite{Carroll2019, Carroll2021, sels2014, TEGMARK2015} 
with the goal to reduce the number of theoretical ingredients needed to
recover classical concepts. 
This is an instructive reconstruction program~\cite{grinbjps} since, as Niels Bohr emphasized, ``it lies in the nature of physical observation that all experience must ultimately be expressed in terms of classical concepts''~\cite[p.~94]{bohr1934}. 

Recovering classical spacetime concepts requires a statement about the meaning of `where' and `when'. In higher-order quantum mechanics~\cite{taranto2025}, one begins with a global Hilbert space not bearing any spatiotemporal labels. The `where' and `when' correspond, then, to the tensor decomposition of this space
into spatial and temporal subsystems. Quantum mereology has shown that the `where' can emerge from spectral data \cite{Loizeau2023, cao2026, Carroll2021}. Whether the `when' can too is the question we address in this paper. If both do, then the spacetime structure would emerge from the algebraic structure alone.

The work on the `where' has focused on deriving locality for Hamiltonian systems: given a generic Hamiltonian (e.g.\ a random matrix under certain constraints), one can find a decomposition into subsystems such that the interactions appear local \cite{Carroll2021, Loizeau2023, Loizeau2024, Zanardi2024, Cotler2019, soulas2025}. This research leaves two open issues. First, the Hamiltonian formalism presupposes a causal structure and a notion of time. While locality can be recovered, causality cannot be derived in such a framework. Second, an approach that begins with choosing a Hamiltonian and a state leaves unanswered fundamental questions about  quantum mechanical observers (agents) and the notion of measurement, which are part of the standard quantum formalism. Whether agents are emergent under certain complexity requirements is a larger topic of investigation in quantum foundations \cite{grin2015,grinbKolm,Carroll2022}. Here, we 
demonstrate that, under certain constraints, quantum mereology gives rise to classical causality. This emergence of the `when' from the algebraic structure alone is linked with the definition of agents.

\subsection{Process matrices and causal structure}

In a standard quantum mechanical setting, consider two parties $A$ and $B$, each operating in a local laboratory with an input Hilbert space $\HAin$ ($\HBin$) and an output Hilbert space $\HAout$ ($\HBout$). 
Each performs a local quantum operation (a quantum instrument) whose outcomes are completely positive maps summing to a trace-preserving one. 
Outcome $i$ of party $A$ is represented by an operator $M_{A_I A_O}^{(i)}$ on $\HAin\otimes\HAout$, and similarly for $B$. 
The most general correlations must be compatible with quantum mechanics and time evolution in each local laboratory, however they must not presuppose
the existence of global time or a global causal structure. These correlations are encoded in a process matrix $W$ on $\Htot = \HAin\otimes\HAout\otimes\HBin\otimes\HBout$ via \cite{Oreshkov2012}
\begin{align}
  p_{ij} = \tr \left(W_{A_I A_O B_I B_O} M_{A_I A_O}^{(i)}M_{B_I B_O}^{(j)}\right).
\label{processmatrix}
\end{align}

Requiring $p_{ij}$ to be a valid probability distribution for every choice of local instruments constrains $W$ to be positive-semidefinite, free of terms that would let a party signal into its own past, and normalized with $\tr W = D_O$, where $D_O$ is the dimension of the output Hilbert space. For qubits, expanding $W$ in Pauli strings turns these conditions into a single statement: only a restricted set $\cP$ of allowed Pauli strings may appear. This set contains all valid processes, while
processes with more restricted causal properties are supported on smaller subsets of $\cP$, as described in
Sec.~\ref{sec:methods}.

A bipartite process is causally ordered  when $B$ cannot signal to $A$ ($\AprecB$). Such processes are supported on a smaller set $\cC\subset\cP$ consisting of all the Pauli strings that do not couple the output of a later agent to the input of an earlier agent. For example, if $\AprecB$, then the Pauli string $\process{\id}{Z}{Z}{\id}$ is in $\cC$, but $\process{Z}{\id}{\id}{Z}$ is not.

Probabilistic mixtures of ordered processes with either $\AprecB$ or $\BprecA$ form the set of causally separable processes.
Among causally non-separable or indefinite processes, some can be represented by putting the mixture of orders in a quantum-controlled superposition \cite{Wechs_2021}. These special cases are superseded by non-causal processes that violate a causal inequality. By analogy with Bell inequalities for non-locality, causal inequalities set bounds that must be satisfied by all causal correlations between parties \cite{Oreshkov2012,Araujo_2015,Branciard_2016,Abbott}. 
For example, in a Guess Your Neighbour's Input game~\cite{GYNIpaper}, the goal is that Alice's input equal Bob's output ($a=y$) and that Bob's input equal Alice's output ($b=x$). For uniformly distributed probabilities and causal correlations, this yields 
$$
P_{GYNI}=P(a=y,b=x)\leq \frac{1}{2},
$$
which can be violated up to a non-classical bound $P_{GYNI}^{max,d=2}\approx 0.5694 > \frac{1}{2}$~\cite{Branciard_2016}. A non-maximal violation is already achieved by the process matrix
\begin{align}
  W_{GYNI} = \frac{1}{4}\!\left(\id
  +\frac{\process{Z}{Z}{Z}{\id}+\process{Z}{\id}{X}{X}}{\sqrt{2}}\right).
\end{align}
Another canonical bipartite non-causal example that lies in $\cP$ outside $\cC$ is the OCB process~\cite{Oreshkov2012}
\begin{align}
  \ocb = \frac{1}{4}\!\left(\id
  +\frac{\process{\id}{Z}{Z}{\id}+\process{Z}{\id}{X}{Z}}{\sqrt{2}}\right).
\end{align} 
Non-causality is not restricted to bipartite systems. Multipartite processes can be causally non-separable even if they are classical and not quantum, the standard example being the tripartite Lugano process where each agent is located simultaneously in the past and in the future of every other agent, yet no causal loops or contradictions occur~\cite{Baumeler2014, Baumeler2016}. Even if non-causal processes cannot be implemented experimentally, causal non-separability can be verified in a device-independent way~\cite{tein2023}. Indefinite causality gave rise to rich experimental and theoretical research \cite{review2026}.

\subsection{Agents, frames, and perspectives}

Causally indefinite processes are defined in (\ref{processmatrix}) using a fixed tensor decomposition of the global Hilbert space $\Htot$ into Alice's and Bob's input and output systems.
We will vary this decomposition and, consequently, redefine agents. Here, agents are constituted by choosing an input and an output, and this choice provides a local time arrow from the input to the output in each agent's laboratory. The agent's operations are local in this time, providing this agent's perspective in the relational approach~\cite{RovRQM,Brukner2020}.

Beyond a local notion of time, operational conceptions of agency in the literature use extra assumptions. For example, one may require that causally indefinite processes be implementable in spacetime, implying that the inputs and outputs correspond to spacetime-localized events \cite{vilasini1,vilasini2}. One may also equip an agent with a reference frame \cite{GuerinBrukner,Kabel2025QuantumCoordinates,apadula2026frameperspectivesprocessmatrices}.
Adding such assumptions in the framework of causally non-separable processes creates conceptual and mathematical tensions \cite{vilasini2,grinbaum2025}. From Alice's perspective, Bob's input is time-delocalized \cite{Oreshkov2018,Wechs_2021,wechs2024}, making impossible any mapping of Bob's input on a well-defined event in Alice's frame. More generally, there are no unitary transformations that switch between agents' perspectives while preserving the time foliation and keeping the global past/future partition fixed~\cite{apadula2026frameperspectivesprocessmatrices}. To avoid these tensions, here we define an agent in a minimal way as a grouping of an input and an output in $\Htot$.

\subsection{Main results}

\begin{figure}
\centering
\includegraphics[width=0.45\textwidth]{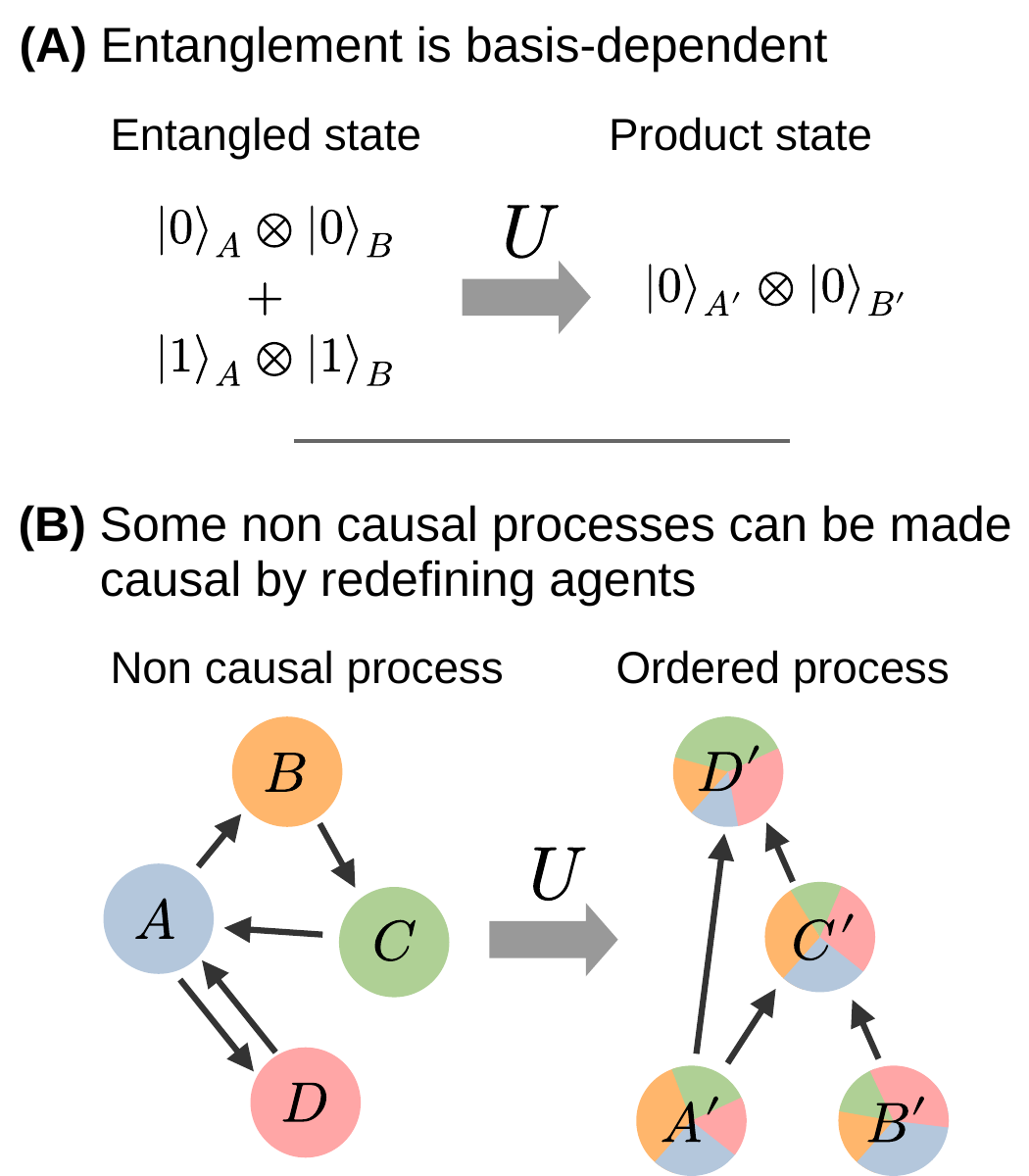}
\caption{\textbf{(A)} Entanglement is basis-dependent, i.e. it depends on the definition of subsystems. Given an entangled pure state $\ket{\psi}$, one can always find a unitary transformation $U$ towards new subsystems on which $\ket{\psi}$ becomes a product state. \textbf{(B)} Similarly, by redefining the decomposition into agents, one can make some non-causal process causal. This redefinition of agents is performed via a unitary $U$ on the global Hilbert space from the old to the new agents' input and output subsystems and time order.
}
\label{fig:drawing2}
\end{figure}

In quantum mechanics, entanglement is not a property of a state alone but depends on the tensor product structure: a state that is entangled under one decomposition factorizes into a product state under another. We ask whether indefinite causality behaves in a similar way, namely whether a change of the input-output subsystems can turn a non-causal process into a causal one (Fig.~\ref{fig:drawing2}).

We study this question by reformulating the problem in terms of spectral conditions. We perform a global unitary transformation in $\Htot$ which decomposes the process in a different way and redefines agents.
A change of subsystems can indeed render certain processes causal, while other processes cannot be ordered under any decomposition. 
Our main result is that, in the thermodynamic limit, randomly chosen quantum processes admit a  decomposition into agents that brings the process close to a causally separable process. This decomposition is preferred in the sense that the associated conception of agents allows one to interpret the process in $\Htot$ as causally definite. The use of the thermodynamic limit hints at the emergence of preferred agency from complexity.

\section{Methods}\label{sec:methods}

We study a bipartite quantum process with 
the total Hilbert space $\Htot = \HAin \otimes \HAout \otimes \HBin \otimes \HBout$.
The central question is: 
\begin{quote}
\emph{Given a process matrix (positive semidefinite trace-$D_O$ operator) $W$ on $\Htot$, does there exist a unitary $U$ such that $W' = UWU^\dagger$ is causally ordered? }
\end{quote}

A valid process matrix is characterized by three conditions: positive semidefiniteness, normalization $\tr W = D_O$, and support on the allowed Pauli set $\cP$, i.e.\ $W$ should lie in the linear span of $\cP$ (for a causally ordered process, in $\cC \subset \cP$) \cite{Oreshkov2012}.
Unitary conjugation preserves the spectrum, hence both positivity and the trace, ensuring that $W' = UWU^\dagger$ is positive semi-definite and correctly normalized. The only condition that might be violated is the support constraint. The problem, therefore, is reduced to finding out whether the unitary orbit of $W$ intersects the subspace of operators supported on $\cC$.

Three cases can be distinguished depending on the structure preserved by $U$:
\begin{itemize}
 \item \textit{Global unitary} ($U$ unrestricted): may mix all four tensor factors $\Ain, \Aout, \Bin, \Bout$.  Since two Hermitian operators are unitarily equivalent if and only if they are isospectral, the problem becomes purely spectral.
 \item \textit{Same-time unitary} ($U_\mathrm{st} = U_{\Ain\Bin} \otimes U_{\Aout\Bout}$): preserves the input/output split, remixes parties within each time slice. 
 \item \textit{Same-party unitary} ($U = U_{\Ain\Aout} \otimes U_{\Bin\Bout}$): preserves the party decomposition, mixes input and output within each party. This corresponds to a local redefinition of the time order.
\end{itemize}

The spectrum of $W$ is the only basis-independent quantity that does not depend on the subsystem decomposition. Therefore, in the global unitary case, the problem reduces to asking what spectra are compatible with causal order. The same-time and same-party cases are more subtle.

Note that $U$ mixes agents $A$ and $B$ and their inputs and outputs. After the transformation, the agents are not the same anymore. For example, the final $A_I'$ might be a mixture of the initial $A_O$ and $B_I$. The unitary $U$ defines a new decomposition into subsystems that are mixtures of old subsystems. The operation of ordering a non-causal process into an ordered process $UWU^\dagger$ has to be interpreted as a redefinition of the subsystems and of the agents, not as an ordering of the initial agents.

\begin{figure}
\centering
\includegraphics[width=0.4\textwidth]{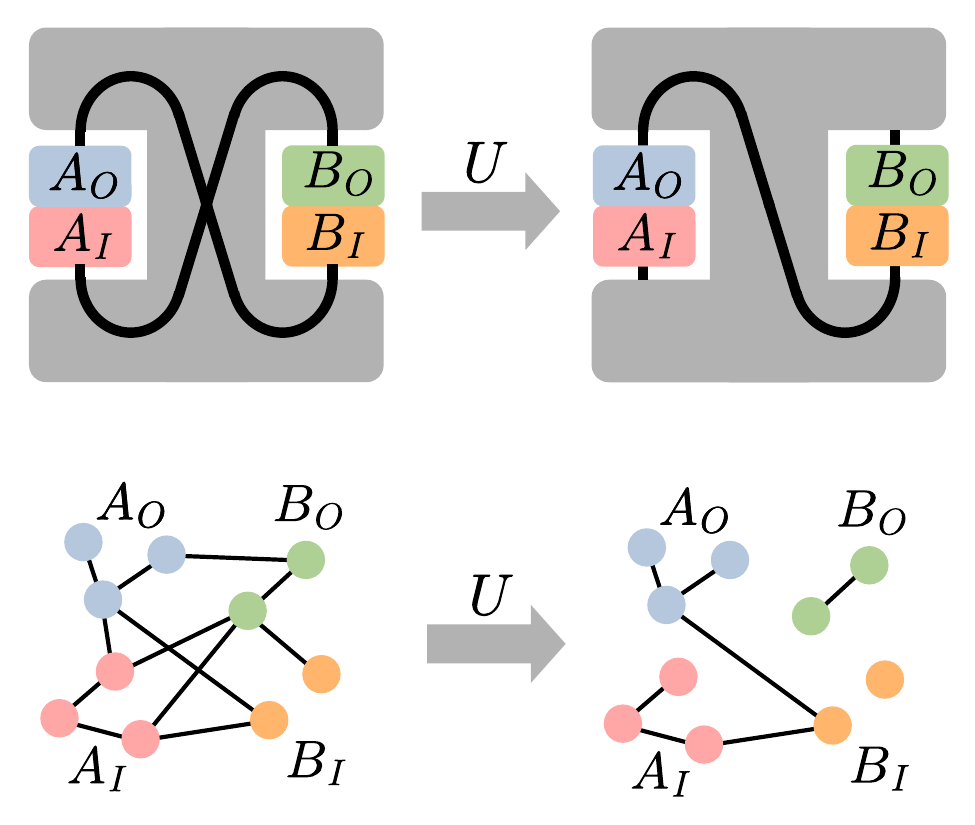}
\caption{\textbf{Top:} Given a process matrix $W$ with indefinite causal order, we look for a unitary transformation $U$ such that $UWU^\dagger$ is causally ordered. \textbf{Bottom:} The above problem can be solved numerically by considering the following many-body problem: given an operator $W$, find an isospectral operator $W'$ supported on a restricted set of Pauli strings $\{P_i\}$\,. Here the nodes represent spins belonging to the different subspaces and the edges represent Pauli strings $\{P_i\}$ that couple these spins.}
\label{fig:drawing}
\end{figure}

Here, we make an attempt to solve these general problems numerically with a focus on the global unrestricted case.
We deploy a method previously used to localize random-matrix Hamiltonians \cite{Loizeau2023} and to partition many-body systems \cite{Loizeau2024,cao2026}, applying it with Pauli string constraints dictated by the causal structure of the process matrix.

If we have a qubit system, it is useful to decompose $W$ in Pauli strings. The set of Pauli strings allowed in $W$ determines the causal ordering \cite{Oreshkov2012}. 
Any operator $W'$ supported on $\cP$ (or $\cC$) can be written as $W'(w) = \sum_{P_k \in \cP} w_k P_k$\,. Finding a unitary $U$ such that $UWU^\dagger \in \cP$ (or $\cC$) is then equivalent to finding coefficients $w$ such that $W'(w)$ has the same spectrum as $W$\, (Figure \ref{fig:drawing}). We solve this by minimizing the spectral distance
\begin{align}
  C(w) = \sum_n \bigl(E_n - \mathcal{E}_n(w)\bigr)^2,
  \label{eq:cost}
\end{align}
where $\{E_n\}$ is the spectrum of $W$ and $\{\mathcal{E}_n(w)\}$ is the spectrum of $W'(w)$. At a zero of $C$ the two operators are isospectral, and the unitary $U$ is recovered by diagonalizing $W$ and $W'$ : if $VEV^\dagger=W$ and $V'\mathcal{E}V'^\dagger=W'$
then $U=V'V^\dagger$, where $E = \operatorname{diag}(\{E_n\})$ and $\mathcal E = \operatorname{diag}(\{\mathcal E_n(w)\})$.

When $W$ can be ordered only approximately, the minimum of $C$ is nonzero and the
recovered operator $\tilde W \equiv U W U^\dagger$ no longer coincides with
$W'$, so it may contain forbidden terms and need not be a valid process. The cost still controls this error. Pauli strings are
orthogonal in the Hilbert--Schmidt inner product, so let $\Pi_{\mathrm{forb}}$
be the orthogonal projector onto the forbidden strings (those outside the
targeted set $\cC$, resp.\ $\cP$). By the Hoffman-Wielandt inequality, the
minimal Frobenius distance between the unitary orbit of $W$ and the fitted
operator $W'(w)$ is the spectral residual,
\begin{align}
  \min_{U}\,\bigl\|\,U W U^\dagger - W'(w)\,\bigr\|^{2}
  = \sum_n \bigl(E_n- \mathcal{E}_n(w)\bigr)^{2}
  = C(w),
\end{align}
attained by $U = V' V^\dagger$. Then $\Pi_{\mathrm{forb}} W' = 0$,
and the forbidden weight of the recovered process is bounded by the cost,
\begin{align}
  \bigl\|\,\Pi_{\mathrm{forb}}\,\tilde W\,\bigr\|
  = \bigl\|\,\Pi_{\mathrm{forb}}(\tilde W - W')\,\bigr\|
  \le \sqrt{C}\,.
\end{align}
At $C = 0$ the bound is saturated trivially, $\tilde W = W'$ exactly. Away from
it, $\sqrt{C}$ quantifies how far the recovered process lies from the causal-order constraints.

Note that this procedure only solves the problem in the case of an unrestricted global unitary. Alternative methods are described in Appendix~\ref{app:numerical} to numerically solve the same-time and same-party unitary versions of the problem.

\section{Results}

\subsection{Examples of causal redefinitions}

First, we provide several examples of unitary transformations of process matrices. For illustration purposes, we use process matrices that are well-known in the literature.

\textbf{Bipartite case.} The OCB process is a process with an indefinite causal order, which violates causal inequalities~\cite{Oreshkov_2016, Branciard_2016}.
However, $\ocb$ can be unitarily mapped to a causally ordered process. There exists a global unitary $U$ such that
\begin{align}
  U\ocb U^\dagger = \frac{\id}{4} + \frac{\process{\id}{Z}{Z}{\id}}{4}
\end{align}
is supported only on $\id$ and $A_OB_I$ Pauli strings and is trivial on $B_O$, placing it in the $\AprecB$ class. This can be checked by comparing spectra: both processes have spectrum $\{0,\,\frac{1}{2}\}$ with multiplicities $8,\,8$.

Causal ordering of $\ocb$ can also be achieved by the more restricted class of same-party unitaries, which preserve the party decomposition and only mix input and output within each party's laboratory. There exists $U=U_{A_IA_O} \otimes U_{B_IB_O}$ with $U_{A_IA_O}=U_{B_IB_O}$ such that
\begin{align}
  U\ocb U^\dagger = \frac{\id}{4} + \frac{1}{4\sqrt{2}}(&\process{\id}{Z}{Z}{\id} \nonumber\\
  + &\process{Z}{\id}{Y}{\id}),
\end{align}
again trivial on $B_O$ and thus $\AprecB$-ordered.
The additional constraint $U_{A_IA_O}=U_{B_IB_O}$ implies that $U$ acts on the time axis in the same way in $A$ and $B$. 
By contrast, causal ordering using a same-time unitary is ruled out for $\ocb$ (Proposition~\ref{prop:sametime_obstruction}).

\textbf{Tripartite case.} A well-known non-causal tripartite process is the so-called Lugano process $W_L$. It has a cumbersome process matrix that we do not reproduce  here~\cite{Araujo2017, Baumeler2016}. There exists a general $U$ such that 
\begin{align}
UW_L U^\dagger
= \frac{1}{8}&\id_{C_IC_O}\Big(
\id_{A_IA_OB_IB_O}
+ \process{Z}{\id}{\id}{\id} \nonumber \\
&+ \process{\id}{Z}{\id}{\id} 
+ \process{Z}{Z}{\id}{\id}\nonumber\\
&+ \process{\id}{\id}{Z}{\id} 
+ \process{Z}{\id}{Z}{\id}\nonumber\\
&+ \process{\id}{Z}{Z}{\id} 
+ \process{Z}{Z}{Z}{\id}
\Big)
\end{align}
is an ordered process that satisfies $\AprecB$ and $A$ and $B$ uncorrelated with $C$. This $U$ is a general unitary that mixes both agents' inputs and outputs and their time ordering.

All these examples can be checked by diagonalizing $W$ before and after transformation. What we learn here is that the causal properties of a process depend on the choice of subsystems: some non-causal processes can be made causal by choosing a different decomposition into subsystems.
In the next section, we formalize and generalize this: what conditions does the spectrum of $W$ need to satisfy to be turned into a causally ordered process with a different tensor factorization?

\subsection{A spectral criterion for order}
\label{subsec:spectral_criterion}
The examples above show that a non-causal process in one decomposition may become causally ordered in another. We now ask when this is possible in general.

We focus on a \emph{total order} between agents. By this we mean that the agents in the new decomposition can be arranged in a sequence
\begin{equation}
\pi(1)\prec\pi(2)\prec\cdots\prec\pi(n)~,
\end{equation}
with a single last agent, $\pi(n)$. The labels here refer to the agents after the subsystem decomposition has been redefined; they need not coincide with the agents in the original description of $W$.

Total order between agents is but one special case of the allowed multipartite causal structures. More generally, the agents' laboratories need not form a single causal sequence: it may be true that, for some pairs of agents, neither one precedes the other; or there may be several last agents, meaning that none of them has a later agent in the causal order
\cite{CostaShrapnel2016,OreshkovGiarmatzi2016,Wechs2019}.
Similarly, when laboratories are embedded in a relativistic spacetime, spacelike-separated laboratories are not ordered relative to one another by the spacetime causal relation
\cite{VilasiniRenner2024}.
We restrict our study to the total order for the benefit of having a unique last agent. The identity factor on the output of this agent will give rise to the spectral degeneracy used below. 

For the $n$-agent setting, write
\begin{equation}
\Htot = \bigotimes_{k=1}^{n} \left( \mathcal H_{k,I}\otimes\mathcal H_{k,O}\right)~,
\label{eq:process_hilbert_space}
\end{equation}
and define
\begin{equation}
D_O\coloneqq\prod_{k=1}^{n}d_{k,O}~, \qquad D\coloneqq\dim\Htot~.
\end{equation}
As before, valid process matrices are positive semidefinite, have trace $D_O$, and satisfy the linear process constraints \cite{Oreshkov2012,Oreshkov_2016,Branciard_2016_witness}.

For a total order $\pi$, let $\mathscr W_\pi$ denote the set of valid process matrices compatible with that order in the new agent decomposition. We define its global-unitary orbit by
\begin{equation}
\mathcal O_\pi \coloneqq \left\{ UXU^\dagger: X\in\mathscr W_\pi~,~ U\in\mathsf U(\Htot) \right\}~.
\label{eq:fixed_order_orbit}
\end{equation}
Thus $W\in\mathcal O_\pi$ means that $W$ need not be ordered in the decomposition in which it was originally written. Rather, it is globally unitarily equivalent to a valid process $X$ that has the order $\pi$ in a redefined decomposition. For normalized operators, we also use
\begin{equation}
\widehat{\mathcal O}_\pi
\coloneqq
\left\{
X/D_O:X\in\mathcal O_\pi
\right\}~.
\label{eq:normalized_fixed_order_orbit}
\end{equation}

Consider first the order
\begin{equation}
1\prec2\prec\cdots\prec n~.
\end{equation}
The last agent is still a local laboratory with an input and an output subsystems, but it has no later laboratory to which it can signal. Note that the last agent is not a deficient agent in the global future with just one input, as in some attempts to ``trade'' causal order for locality~\cite{Kunjwal_2023,Dourdent2025,grinbaum2025}. Being last  means only that the agent's output cannot influence any other agent.
Its output is therefore an open final wire, and the fixed-order constraints \cite{Branciard_2016_witness} imply
\begin{equation}
X\in\mathscr W_{1\prec\cdots\prec n} \quad \Longrightarrow \quad X= \frac{\mathbb 1_{n,O}}{d_{n,O}} \otimes K~,
\label{eq:last_output_factorisation_main}
\end{equation}
where $K\geq0$ acts on all factors except $\mathcal H_{n,O}$. The remaining fixed-order constraints act recursively on $K$. We recall this factorization in Appendix~\ref{app:fixed_order_factorization}. 

This factorization has an immediate spectral consequence. If $K$ has an eigenvalue $\mu$, then $X$ has the eigenvalue $\mu/d_{n,O}$ repeated $d_{n,O}$ times. Every distinct eigenvalue of a fixed-order process must therefore have multiplicity divisible by the output dimension of the last agent.

\begin{proposition}[Spectral obstruction to exact fixed order]
\label{prop:spectral_obstruction}
Let $X\in\mathscr W_{1\prec\cdots\prec n}$. Then every distinct eigenvalue of $X$ has multiplicity divisible by $d_{n,O}$. Consequently, if $d_{n,O}>1$, an operator $W\geq0$ with $\tr W=D_O$ and simple spectrum cannot belong to $\mathcal O_{1\prec\cdots\prec n}$.
\end{proposition}

The proposition gives a necessary condition. The converse requires more work. The correct eigenvalue multiplicities do not by themselves make an operator a valid process: one must also choose eigenvectors that satisfy the linear fixed-order constraints. Under a mild condition on the subsystem dimensions, such an eigenbasis can always be constructed. The spectral degeneracy is then the only obstruction.

The required dimension condition is
\begin{equation}
d_{k,O}\mid d_{k+1,I}~,
\qquad
k=1,\dots,n-1~.
\label{eq:chain_divisibility}
\end{equation}
Here $a\mid b$ means that $a$ divides $b$. The condition is automatic when the elementary agents have equal input and output dimensions.

This condition allows the next input space $\mathcal H_{k+1,I}$ to be decomposed into blocks of dimension $d_{k,O}$. The output $O_k$ can then be paired with each such block in a generalized Bell basis. Every basis state has a maximally mixed marginal on $O_k$, so operators diagonal in this basis satisfy the recursive fixed-order constraints.

The construction is given in Appendix~\ref{app:sufficiency_spectral_condition}. There we build the generalized Bell basis explicitly, assign the required reduced eigenvalues to its basis vectors, and finally attach the open-output factor $\mathbb 1_{n,O}/d_{n,O}$. The result is a valid process in $\mathscr W_{1\prec\cdots\prec n}$ with the prescribed spectrum.

\begin{theorem}[Exact spectral characterization]
\label{thm:exact_characterization}
Assume condition~\eqref{eq:chain_divisibility}. Let $W\geq0$ on $\Htot$ with $\tr W=D_O$. Let $\lambda_\alpha$ denote the distinct eigenvalues of $W$, with multiplicities $m_\alpha$. Then
\begin{equation}
W\in\mathcal O_{1\prec\cdots\prec n}
\quad\Longleftrightarrow\quad
d_{n,O}\mid m_\alpha
\quad\text{for all }\alpha~.
\label{eq:exact_characterization}
\end{equation}
For a general total order $\pi$, the relevant divisor is the output dimension of the last agent, $d_{\pi(n),O}$.
\end{theorem}

The theorem gives an exact spectral test for compatibility with a chosen total order. If the multiplicities fail the divisibility condition, no choice of eigenvectors and no global redefinition of the agents can produce that order. If the condition holds, an eigenbasis can be chosen for which all fixed-order process constraints are satisfied. For example, if the output of the last agent is a qubit, every eigenvalue must occur with even multiplicity. The spectrum must therefore have the form 
\begin{equation} 
(\lambda_1,\lambda_1,\lambda_2,\lambda_2,\ldots)~. 
\end{equation} 
By contrast, an operator drawn from an absolutely continuous finite-dimensional ensemble has a simple spectrum with probability one \cite{Mehta2004,Forrester2010}, 
\begin{equation} 
\lambda_1>\lambda_2>\lambda_3>\cdots~. 
\end{equation} 
Such an operator may satisfy positivity and the trace normalization, but these are only two of the requirements for a valid process. A total order imposes additional linear constraints \cite{Oreshkov2012,Oreshkov_2016,Branciard_2016_witness}, including the identity factor on the final output, and therefore requires the corresponding spectral degeneracies. Since a global unitary cannot change eigenvalue multiplicities, an operator with simple spectrum cannot be transformed into a total-order process with a nontrivial final output.

\subsection{Changing the number of agents}
\label{subsec:changing_agents}

The spectral criterion also constrains how the global Hilbert space may be decomposed into different numbers of agents. The agents considered in this section belong to the new decomposition in which a strict total order is sought, and are denoted by tilded symbols such as $\widetilde A$, $\widetilde B$, and $\widetilde C$.

Suppose that the global Hilbert space is built from $N$ elementary input-output pairs labelled by $1,\ldots,N$. A decomposition into $m$ agents is specified by a partition
\begin{equation}
\{1,\ldots,N\} = G_1\sqcup G_2\sqcup\cdots\sqcup G_m~.
\label{eq:agent_partition}
\end{equation}
Each set $G_a$ defines one agent $\widetilde A_a$, with input and output spaces
\begin{equation}
\begin{aligned}
\mathcal H_{\widetilde A_a,I} &= \bigotimes_{k\in G_a}\mathcal H_{k,I}~,\\
\mathcal H_{\widetilde A_a,O} &= \bigotimes_{k\in G_a}\mathcal H_{k,O}~.
\end{aligned}
\label{eq:grouped_agent_spaces}
\end{equation}
We then ask whether $W$ can be globally unitarily transformed into a valid process with the total order
\begin{equation}
\widetilde A_1 \prec \widetilde A_2 \prec\cdots\prec \widetilde A_m~.
\end{equation}

By \emph{splitting an agent}, we mean dividing  $G_a$ into two or more disjoint subsets  assigned to separate agents who are sequential in the total order. Conversely, \emph{merging agents} means replacing two or more consecutive agents by a single agent who is assigned all their inputs and outputs. Both splitting and merging change the number of agents, but not the dimension of the global Hilbert space. A detailed discussion of splitting and merging agents is given in Appendix~\ref{app:agent_redefinition}.

Let the distinct eigenvalues of $W$ have multiplicities $m_1,\ldots,m_r$, and define
\begin{equation}
g(W) \coloneqq \gcd(m_1,\ldots,m_r)~.
\label{eq:gcd_multiplicity}
\end{equation}
Thus $g(W)$ is the largest integer that divides every eigenvalue multiplicity. If $\widetilde L=\widetilde A_m$ is the last agent in the proposed order, then
\begin{equation}
d_{\widetilde L,O} = \prod_{k\in G_m}d_{k,O}~.
\end{equation}
Provided that the dimension condition analogous to Eq.~\eqref{eq:chain_divisibility} holds for this decomposition, Theorem~\ref{thm:exact_characterization} gives
\begin{equation}
W\in \mathcal O_{\widetilde A_1\prec\cdots\prec\widetilde A_m} \quad\Longleftrightarrow\quad d_{\widetilde L,O}\mid g(W)~.
\label{eq:final_output_rule}
\end{equation}

Equation~\eqref{eq:final_output_rule} determines whether $W$ can be globally unitarily transformed into a valid process with the proposed total order. The multiplicities do not fix a unique number of agents or a unique decomposition. They constrain only the output dimension of the last agent. Splitting the last agent can reduce this dimension and weaken the required spectral degeneracy, whereas merging agents into the last agent has the opposite effect. Splitting or merging agents elsewhere in the order does not change the final-output divisor directly, although it may affect the dimension condition~\eqref{eq:chain_divisibility} needed for sufficiency.

\paragraph{Splitting the last agent.}

Consider three elementary qubit input-output pairs and suppose that
\begin{equation}
g(W)=2~.
\end{equation}
First choose a two-agent decomposition,
\begin{equation}
G_1=\{1\}~, \qquad G_2=\{2,3\}~,
\end{equation}
and consider the order
\begin{equation}
\widetilde A\prec\widetilde B~.
\end{equation}
The last agent $\widetilde B$ contains two elementary output qubits, and therefore
\begin{equation}
d_{\widetilde B,O}=2^2=4~.
\end{equation}
The exact spectral condition would require
\begin{equation}
4\mid g(W)~.
\end{equation}
Since $g(W)=2$, no global unitary can transform $W$ into a valid process with this two-agent order.

Now split the last agent by replacing $G_2=\{2,3\}$ with two agents,
\begin{equation}
G_1=\{1\}~, \qquad G_2=\{2\}~, \qquad G_3=\{3\}~,
\end{equation}
and consider
\begin{equation}
\widetilde A\prec\widetilde B\prec\widetilde C~.
\end{equation}
The last agent $\widetilde C$ now has a single output qubit, so
\begin{equation}
d_{\widetilde C,O}=2~.
\end{equation}
Since $2\mid g(W)$, the same spectrum satisfies the exact condition for this three-agent order. Under the dimension assumptions of Theorem~\ref{thm:exact_characterization}, there therefore exists a global unitary that transforms $W$ into a valid process with the order
$\widetilde A\prec\widetilde B\prec\widetilde C$.

\paragraph{Merging agents.}

The reverse operation has the opposite effect. For $g(W)=2$, the three-agent order
\begin{equation}
\{1\}\prec\{2\}\prec\{3\}
\end{equation}
satisfies the spectral condition because the last output has dimension $2$. If the agents containing the elementary input-output pairs $2$ and $3$ are merged, the proposed order becomes
\begin{equation}
\{1\}\prec\{2,3\}~.
\end{equation}
The output dimension of the last agent then increases from $2$ to $4$, and the same spectrum fails the exact condition. Splitting or merging the last agent can therefore change whether $W$ can be transformed into a process with a given strict total order.

For equal elementary input and output dimensions,
\begin{equation}
d_{k,I}=d_{k,O}=d~,
\end{equation}
the spectral constraint can be summarized by the integer
\begin{equation}
\ell_d(W) \coloneqq \max \left\{ \ell\geq0:d^\ell\mid g(W) \right\}~.
\label{eq:divisibility_depth}
\end{equation}
If the last agent contains $b$ elementary input-output pairs, then its output dimension is $d^b$, and Eq.~\eqref{eq:final_output_rule} becomes
\begin{equation}
b\leq\ell_d(W)~.
\label{eq:final_agent_depth_rule}
\end{equation}
Thus $\ell_d(W)$ is the largest number of elementary output factors that can be assigned to the last agent without violating the exact spectral condition. It does not determine the rest of the decomposition, and the chain-divisibility condition must still be checked. In particular, $\ell_d(W)=0$ rules out any last agent with a nontrivial elementary output, while $\ell_d(W)=1$ or $2$ allows the last agent to contain at most one or two elementary input-output pairs, respectively.

\subsection{Restricted redefinitions of subsystems}
\label{subsec:restricted_subsystem_changes}

The spectral criterion above relies on allowing an arbitrary global unitary on $\Htot$. Such a transformation may mix input and output systems, and can therefore redefine both the agents and the local input-output structure assigned to them. If the allowed transformations preserve part of the original subsystem structure, the spectrum alone is no longer sufficient.

A useful intermediate case is a same-time unitary in the bipartite setting,
\begin{equation}
U_{\rm st}
=
U_{A_I B_I}\otimes U_{A_O B_O}~.
\label{eq:sametime_unitary}
\end{equation}
This transformation preserves the global input-output split. It may redefine how the total input space is divided between the two agents, and independently how the total output space is divided between them, but it cannot mix an input system with an output system.

Let
\begin{equation}
\mathcal H_I
=
\mathcal H_{A_I}\otimes\mathcal H_{B_I}~,
\qquad
\mathcal H_O
=
\mathcal H_{A_O}\otimes\mathcal H_{B_O}~,
\end{equation}
and let $d_I=\dim\mathcal H_I$. Choose a Hilbert--Schmidt orthonormal basis $\{A_k\}_{k=1}^{d_I^2-1}$ of traceless Hermitian operators on $\mathcal H_I$. Any Hermitian operator $W$ can then be written as
\begin{equation}
W = \frac{\mathbb 1_I}{d_I}\otimes C_0 + \sum_{k=1}^{d_I^2-1}A_k\otimes B_k~,
\label{eq:input_output_schmidt}
\end{equation}
where
\begin{equation}
C_0=\tr_I W~, \qquad B_k = \tr_I\!\left[ (A_k\otimes\mathbb 1_O)W \right]~.
\end{equation}
The output operators associated with the nontrivial input dependence span the subspace
\begin{equation}
\mathcal S_{\rm out}(W) \coloneqq \operatorname{span} \left\{ B_k:1\leq k\leq d_I^2-1 \right\} \subset \mathcal B(\mathcal H_O)~.
\label{eq:output_operator_system}
\end{equation}
Although the individual operators $B_k$ depend on the chosen basis of traceless input operators, their span does not.

Under the same-time transformation~\eqref{eq:sametime_unitary}, this subspace changes only by conjugation on the total output space:
\begin{equation}
\mathcal S_{\rm out} \left( U_{\rm st}WU_{\rm st}^\dagger \right) = U_{A_OB_O}\, \mathcal S_{\rm out}(W)\, U_{A_OB_O}^\dagger~. \label{eq:output_subspace_transformation}
\end{equation}
Its dimension and its internal algebraic relations, such as commutativity, are therefore preserved.

Suppose that a same-time unitary transforms $W$ into a valid process with the strict order $\widetilde A\prec\widetilde B$. Since the output of the last agent $\widetilde B$ must be trivial, the transformed output subspace must satisfy
\begin{equation}
U_{A_OB_O}\, \mathcal S_{\rm out}(W)\, U_{A_OB_O}^\dagger \subset \mathcal B(\mathcal H_{\widetilde A_O}) \otimes \mathbb 1_{\widetilde B_O}~. \label{eq:output_subspace_A_before_B}
\end{equation}
For the opposite order $\widetilde B\prec\widetilde A$, it must instead satisfy
\begin{equation}
U_{A_OB_O}\, \mathcal S_{\rm out}(W)\, U_{A_OB_O}^\dagger \subset \mathbb 1_{\widetilde A_O} \otimes \mathcal B(\mathcal H_{\widetilde B_O})~.
\label{eq:output_subspace_B_before_A}
\end{equation}
The restricted problem is therefore stronger than the spectral one. It asks whether the output subspace can be moved entirely into the output algebra of the earlier agent while the global input-output split remains fixed.

\begin{proposition}[Same-time obstruction]
\label{prop:sametime_obstruction}
Suppose that $\mathcal S_{\rm out}(W)$ contains a $q$-dimensional real subspace of mutually commuting traceless Hermitian operators. If
\begin{equation}
q> \max \left( d_{A_O}-1,d_{B_O}-1 \right)~,
\label{eq:sametime_bound}
\end{equation}
then no same-time unitary can transform $W$ into a valid process with either strict bipartite order.
\end{proposition}

Indeed, a commuting family of Hermitian operators on a $d$-dimensional Hilbert space can be simultaneously diagonalized, and its traceless part has dimension at most $d-1$. The bound in Proposition~\ref{prop:sametime_obstruction} is therefore preserved under output conjugation but is too large to fit inside the output algebra of either possible earlier agent.

For two-qubit outputs, the largest commuting traceless subspace contained in either single-agent output algebra has dimension one. Two linearly independent commuting output observables are therefore already enough to obstruct both strict orders under a same-time unitary. This accounts for the distinction found in the OCB example: the process can be transformed into a causally ordered one by a transformation that mixes input and output systems, but not by a transformation that only redefines the parties separately within the input and output spaces. The proof is given in Appendix~\ref{app:same_time_obstruction}.

\subsection{Distance to a total order at large dimension}
\label{subsec:large_size_numerics}

The exact spectral characterization in
Theorem~\ref{thm:exact_characterization} is rigid. If the output of the last agent is nontrivial, an operator can be globally unitarily equivalent to a valid process with a given total order only if its spectrum has exact degeneracies. A spectrum drawn from a continuous finite-dimensional random-matrix ensemble almost surely has no such degeneracies \cite{Forrester2010,Mehta2004}. We therefore ask whether this can be achieved  approximately: how close is a typical high-dimensional spectrum to one realized by a valid process with a fixed total causal order?

We sample normalized positive operators from the Wishart ensemble,
\begin{equation}
\rho = \frac{GG^\dagger}{\tr(GG^\dagger)}~,
\end{equation}
where the entries of $G$ are independent complex Gaussian random variables. The Wishart draw is used here only to supply a random spectrum. It need not itself satisfy the process-matrix constraints in the decomposition in which it is sampled.

For each spectrum, we fix a total order $\pi$ and minimize the cost function in Eq.~\eqref{eq:cost} over all valid process matrices with that order. Thus the minimization is not performed with respect to one particular ordered process, but over the entire fixed-order set $\mathscr W_\pi$\,. A vanishing minimum means that the sampled spectrum is realized exactly by some process in $\mathscr W_\pi$\,. A nonzero minimum measures the remaining spectral mismatch.

\begin{figure}
\centering
\includegraphics[width=0.49\textwidth]{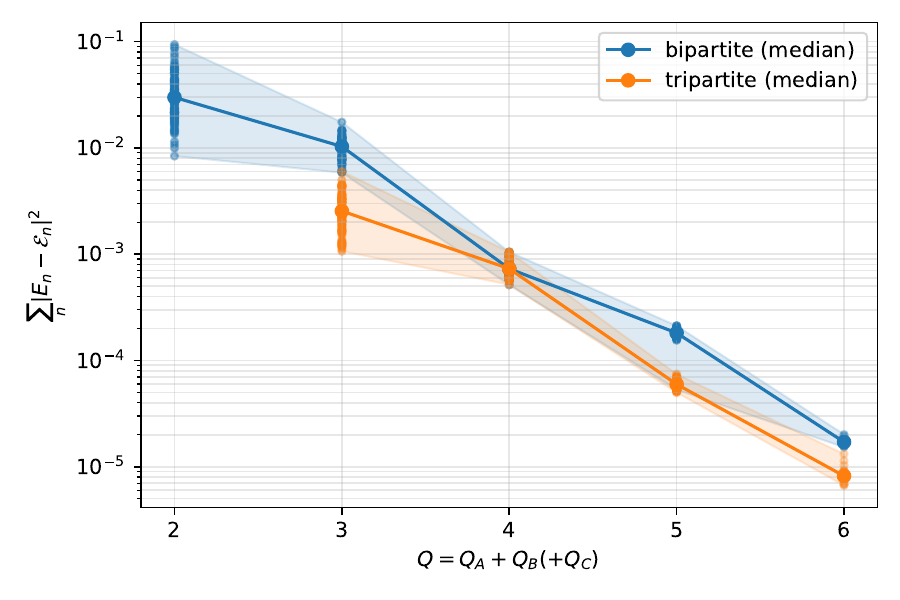}
\caption{Minimum of the spectral cost in Eq.~\eqref{eq:cost} for spectra sampled from the Wishart ensemble. We draw $\rho=GG^\dagger/\tr(GG^\dagger)$ and, for a fixed total causal order $\pi$, minimize over all valid process matrices in $\mathscr W_\pi$. The total Hilbert-space dimension is $D=2^Q$, where $Q$ is the total number of elementary qubits. The residual cost decreases rapidly with $Q$, showing that random high-dimensional spectra approach the set of spectra realized by valid processes with the fixed total order. Each point is averaged over $100$ independent realizations.}
\label{fig:cost}
\end{figure}

In the qubit simulations shown in Fig.~\ref{fig:cost}, we consider a fixed bipartite order and a fixed tripartite order. Let $Q$ denote the total number of elementary qubits in $\Htot$, so that
\begin{equation}
D=2^Q~.
\end{equation}
We increase $Q$ while keeping the input and output dimensions of the different agents as balanced as possible.

The residual cost decreases rapidly with $Q$ in both cases. This does not mean that a Wishart spectrum satisfies the exact multiplicity condition at finite dimension: exact degeneracies remain absent almost surely. Rather, its distance to the set of spectra realized by valid processes with the fixed total order becomes small as the Hilbert-space dimension grows.

\subsection{Why the large-dimension approximation works}
\label{subsec:approximate_orderability}

The decrease of the numerical cost has a simple explanation. For a fixed total order, Theorem~\ref{thm:exact_characterization} says that exact membership in the corresponding global-unitary orbit is a spectral condition: the eigenvalues must come in degeneracy blocks fixed by the output dimension of the last agent. Thus the approximation problem is to ask how far a given spectrum is from one with this block-degeneracy pattern.

Fix a total order $\pi$ for which Theorem~\ref{thm:exact_characterization} applies, and write $d_{\pi(n),O}$ for the output dimension of the last agent. For a normalized operator $\rho=W/D_O$\,, define the Hilbert--Schmidt distance to the normalized total-order orbit by
\begin{equation}
\widehat\delta^{(2)}_\pi(\rho) \coloneqq \inf_{\sigma\in\widehat{\mathcal O}_\pi} \|\rho-\sigma\|_2~.
\label{eq:hs_distance}
\end{equation}
Let
\begin{equation}
\nu_1\geq\nu_2\geq\cdots\geq\nu_D\geq0~, \qquad \sum_{i=1}^{D}\nu_i=1~,
\label{eq:eigenvalue_vector}
\end{equation}
be the eigenvalues of $\rho$\,, where $D=\dim\Htot$\,. Exact compatibility with the order $\pi$ requires this ordered list to be constant in blocks of length $d_{\pi(n),O}$.

Assume that $d_{\pi(n),O}$ divides $D$, as it does for the process decompositions considered here, and set
\begin{equation}
N_\pi \coloneqq \frac{D}{d_{\pi(n),O}}~.
\end{equation}
For $a=1,\ldots,N_\pi$\,, define
\begin{equation}
B_a \coloneqq \left\{(a-1)d_{\pi(n),O}+1,\ldots,a d_{\pi(n),O}\right\}
\label{eq:spectral_blocks}
\end{equation}
and
\begin{equation}
\bar\nu_a \coloneqq \frac{1}{d_{\pi(n),O}} \sum_{j\in B_a}\nu_j~.
\label{eq:block_average}
\end{equation}
The closest block-degenerate spectrum is obtained by replacing the eigenvalues in each block $B_a$ by their average $\bar\nu_a$.

\begin{theorem}[Distance to a total-order orbit]
\label{thm:distance_formula}
Under the assumptions of Theorem~\ref{thm:exact_characterization},
\begin{equation}
\begin{aligned}
\widehat\delta^{(2)}_\pi(\rho)^2 &= \sum_{a=1}^{N_\pi} \sum_{j\in B_a} (\nu_j-\bar\nu_a)^2\\
&= \tr(\rho^2) - d_{\pi(n),O} \sum_{a=1}^{N_\pi} \bar\nu_a^2~.
\end{aligned}
\label{eq:distance_formula}
\end{equation}
In particular,
\begin{equation}
\widehat\delta^{(2)}_\pi(\rho)^2 \leq \tr(\rho^2)~.
\label{eq:purity_bound}
\end{equation}
\end{theorem}

Equation~\eqref{eq:distance_formula} says that the distance to the total-order orbit is the variance of the spectrum inside the blocks selected by the order. The relevant dimensional data are therefore not only the total dimension $D$. The last agent's output dimension fixes the block size, while the remaining agents' input and output dimensions must satisfy the chain-divisibility condition~\eqref{eq:chain_divisibility} for Theorem~\ref{thm:exact_characterization} to apply.

There is, however, no single way to make the process larger. The same increase of $D$ may come from adding more agents, from enlarging the input and output spaces of fixed agents, from increasing only input spaces or only output spaces, or from assigning the Hilbert-space dimension very unevenly among the agents. These choices describe different limits. Adding agents lengthens the causal chain. Enlarging fixed agents keeps the same chain but gives the laboratories larger quantum systems. Enlarging output spaces is especially relevant here, because the final output dimension sets the required block size.

For Eq.~\eqref{eq:distance_formula}, however, these differences matter only through the blocks they produce. Once $d_{\pi(n),O}$, $N_\pi$, and the divisibility conditions are fixed, the distance no longer depends on whether the large Hilbert space came from more agents, larger agents, or a different distribution of input and output dimensions. What remains is a spectral question: how much do the eigenvalues vary inside the required blocks?

The purity bound gives a simple sufficient answer. For any sequence of normalized operators $\rho_D$ satisfying the assumptions of Theorem~\ref{thm:exact_characterization},
\begin{equation}
\tr(\rho_D^2)\longrightarrow0 \qquad \Longrightarrow \qquad \widehat\delta^{(2)}_\pi(\rho_D) \longrightarrow0~.
\label{eq:vanishing_purity_condition}
\end{equation}
Thus a large-dimension limit can approach the total-order orbit when two conditions are met: the dimensions grow in a way compatible with the required divisibility conditions, and the spectral weight spreads over an increasing number of eigenvalues. The first condition fixes the blocks. The second makes the eigenvalues nearly constant within them.

The second condition is not automatic. To see this, consider, for each value of $D$\,, the total-order orbit $\widehat{\mathcal O}_{\pi_D}$ satisfying the assumptions of Theorem~\ref{thm:exact_characterization}, and write $d_{\pi_D(n),O}$ for the output dimension of the last agent. Assume that this output is
nontrivial, so that $d_{\pi_D(n),O}>1$\,. Now consider the spiked family
\begin{equation}
\rho_D^{(p)} = p\ket{\psi}\bra{\psi} + (1-p)\frac{\mathbb 1}{D}~, \qquad 0<p<1~.
\label{eq:spiked_family}
\end{equation}
Its eigenvalues are
\begin{equation}
p+\frac{1-p}{D}~, \qquad \underbrace{\frac{1-p}{D}~,\ldots,\frac{1-p}{D}}_{D-1\text{ times}}~.
\end{equation}
The orbit $\widehat{\mathcal O}_{\pi_D}$ requires degeneracy blocks of size $d_{\pi_D(n),O}$\,. The first block therefore contains the spectral spike and $d_{\pi_D(n),O}-1$ small eigenvalues. Its average is
\begin{equation}
\frac{1-p}{D}+\frac{p}{d_{\pi_D(n),O}}~,
\end{equation}
while all later blocks are already constant. Hence Eq.~\eqref{eq:distance_formula} gives
\begin{equation}
\widehat\delta^{(2)}_{\pi_D}(\rho_D^{(p)})^2 = p^2\left(1-\frac{1}{d_{\pi_D(n),O}}\right)~.
\label{eq:spiked_hs_distance}
\end{equation}
Since $d_{\pi_D(n),O}>1$\,, this distance does not vanish as $D$ grows. The additional dimensions only add small eigenvalues around the spike; they do not remove the spike itself.

Thus the approximation is not a consequence of dimension alone. Different large-dimension limits may have different interpretations and may lead to different block structures. Once the block structure is fixed, however, the distance to the total-order orbit is controlled by the spectrum: it decreases when the spectral weight spreads over many eigenvalues, and it need not decrease when a finite part of the weight remains concentrated. The random-matrix calculation below is a controlled way of testing the favourable case, where the dimension grows and the spectrum becomes increasingly spread out.

\subsection{Random spectra and typicality}
\label{subsec:wishart_typicality}
We now apply the distance formula to a standard random-matrix ensemble. The Wishart ensemble is used only as an ensemble of spectra. A matrix sampled from it need not satisfy the process constraints in the decomposition in which it is drawn. The question is instead whether its spectrum lies close to one realized by a valid process with total causal order. This is sufficient because the distance in Eq.~\eqref{eq:hs_distance} depends only on the eigenvalues.

Let
\begin{equation}
G\in\mathbb C^{D\times D}~, \qquad \rho_D = \frac{GG^\dagger}{\tr(GG^\dagger)}~,
\label{eq:balanced_wishart}
\end{equation}
where the entries of $G$ are independent complex Gaussian random variables. This defines the balanced Hilbert--Schmidt induced ensemble. Equivalently, $\rho_D$ is obtained by tracing out a $D$-dimensional subsystem from a Haar-random pure state on
$\mathbb C^D\otimes\mathbb C^D$
\cite{Zyczkowski_2001,Mehta2004,Forrester2010}. Its average purity is
\begin{equation}
\mathbb E\,\tr(\rho_D^2) = \frac{2D}{D^2+1}~.
\label{eq:wishart_purity}
\end{equation}

At every finite $D$\,, the spectrum of $\rho_D$ is simple almost surely. Indeed, repeated eigenvalues form a set of measure zero in an absolutely continuous matrix ensemble
\cite{Mehta2004,Forrester2010}. Therefore, whenever
$d_{\pi(n),O}>1$\,, a Wishart spectrum almost surely fails the exact multiplicity condition in Theorem~\ref{thm:exact_characterization}. It cannot lie exactly in the total-order orbit $\widehat{\mathcal O}_\pi$.

Its distance to that orbit behaves differently. Combining the purity bound in Eq.~\eqref{eq:purity_bound} with Eq.~\eqref{eq:wishart_purity} gives
\begin{equation}
\mathbb E\!\left[ \widehat\delta^{(2)}_\pi(\rho_D)^2 \right] \leq \frac{2D}{D^2+1}~.
\label{eq:wishart_distance_bound}
\end{equation}
The right-hand side vanishes as $D\to\infty$. Thus, although a Wishart spectrum almost surely fails the exact degeneracy condition at every finite dimension, its mean squared distance to the total-order orbit tends to zero.

If $D=2^Q$, where $Q$ is the total number of elementary qubits, then
\begin{equation}
\mathbb E\!\left[ \widehat\delta^{(2)}_\pi(\rho_D)^2 \right] = O(2^{-Q})~.
\end{equation}
The expected squared distance therefore decreases exponentially with $Q$, while the root-mean-square distance is bounded by a quantity of order $2^{-Q/2}$.

The expectation bound also implies convergence in probability. For every $\varepsilon>0$, Markov's inequality gives
\begin{equation}
\mathbb P\!\left( \widehat\delta^{(2)}_\pi(\rho_D)>\varepsilon \right) \leq \frac{1}{\varepsilon^2} \frac{2D}{D^2+1} \xrightarrow[D\to\infty]{}0~.
\label{eq:wishart_probability_bound}
\end{equation}
Equivalently,
\begin{equation}
\widehat\delta^{(2)}_\pi(\rho_D) \xrightarrow[D\to\infty]{\mathbb P} 0~.
\label{eq:wishart_convergence}
\end{equation}
Hence, for any fixed tolerance $\varepsilon$, the probability that a Wishart spectrum lies farther than $\varepsilon$ from the total-order orbit tends to zero as the Hilbert-space dimension grows.

This explains the numerical decay observed in Fig.~\ref{fig:cost}. In the large-dimension limit considered here, the number of agents and their total ordering are fixed, while the local input and output dimensions are increased. Exact membership in the total-order orbit remains absent almost surely at every finite $D$, but the distance to that orbit becomes arbitrarily small with high probability.

The mechanism is the spreading of spectral weight. Since
$\mathbb E\,\tr(\rho_D^2)\sim 2/D$, the weight is distributed over an increasing number of eigenvalues, each of which is typically small. Replacing neighboring eigenvalues by their block averages therefore changes the spectrum only slightly in Hilbert--Schmidt norm. Wishart spectra do not become exactly degenerate; rather, they approach spectra that can be realized by valid processes with the chosen total causal order after a suitable redefinition of the agents.

The same argument applies to the more general induced ensemble
\begin{equation}
G\in\mathbb C^{D\times s}~.
\end{equation}
In this case,
\begin{equation}
\mathbb E\,\tr(\rho^2) = \frac{D+s}{Ds+1}~.
\label{eq:wishart_general_purity}
\end{equation}
Whenever $\min(D,s)\to\infty$, the average purity tends to zero, and the same purity bound gives
\begin{equation}
\widehat\delta^{(2)}_\pi(\rho) \xrightarrow{\mathbb P}0~.
\end{equation}
The balanced choice $s=D$ is the simplest example of this regime.

\section{Discussion}

Every quantum object: a Hamiltonian, a density matrix, a state, and a process matrix, has physical meaning only relative to a tensor-product decomposition of the global Hilbert space into subsystems. For a process matrix, this decomposition implies the grouping of an input and an output to define agents. We have shown that causal definiteness is not intrinsic to the process matrix alone, but depends on the definition of agents. Under arbitrary global unitary transformations, agents and subsystems change, while spectral data remain invariant. The spectrum encodes causal constraints: to be totally ordered, a process must have specific spectral degeneracies. Generic finite-dimensional spectra, which we studied on random process matrices, do not meet this condition.

In the thermodynamic limit, however, typical high-dimensional process matrices have flat random spectra, and this flatness is precisely what makes them close to causally ordered processes. Smoothness is often associated with quantum chaos: chaotic spectra have a smooth density of states and level repulsion \cite{DAlessio2016, Mehta2004}, and once the process matrix is normalized, the typical level spacing vanishes as the dimension grows. The mechanism isolated here is spectral flatness, or equivalently low purity: eigenvalues can be grouped into almost-degenerate blocks with small error. The degeneracies that exact order would require are then recovered approximately. For a total Hilbert-space dimension $D=2^Q$, our Wishart estimate gives a squared spectral distance of order $O(2^{-Q})$, matching the rapid decay seen in the numerics. Previously an analogous phenomenon has been reported in quantum mereology, where large chaotic systems were easier to localize than small ones~\cite{Loizeau2023}; there too, the underlying reason was typicality in a large Hilbert space.

This also gives a useful entropy interpretation. High-purity, low-effective-rank spectra can obstruct approximate orderability, because a sharp spectral spike cannot be hidden by changing the tensor product structure. By contrast, flatter spectra have a larger effective degeneracy: many neighboring eigenvalues can be grouped with little cost. In this sense, high-dimensional random spectra do not become exactly totally ordered. Rather, they become so flat that they are effectively compatible with a totally ordered process after a global redefinition of the agent decomposition.

Two limitations are worth recalling to avoid confusion. First, our result is about approximate total order, while the set of processes with exact total order remains of measure zero. Second, unitary transformations on the global Hilbert space will in general redefine the agents, mixing inputs with outputs across parties. The emergent causal order applies to the new subsystems and should not be confused with an ordering of the original agents. Also, once the allowed transformations are restricted, the spectrum no longer tells the whole story. The same-time obstruction above is an example: if the unitary is required to preserve the input-output split, then the output operator system becomes an additional invariant.

One motivation for our work was the idea that, within quantum gravity, space and time may emerge from abstract data. This idea is e.g.\ central within the AdS/CFT framework \cite{Maldacena:1997re} where the RT-formula \cite{Ryu:2006bv} (or an iteration thereof that applies to so-called quantum extremal surfaces \cite{Engelhardt2015}) shows how geometry in the semi-classical limit of quantum gravity can be reconstructed from the dynamics and entanglement-correlations of a conformal field theory (the ``boundary theory''), which is considered the fundamental theory underlying that semi-classical limit. Note, however, that this supposed CFT-geometry duality already starts with assumed dynamics and thus an assumed causality structure in the boundary theory. The process matrix framework may be a step towards a more general emergence paradigm, that assumes no \emph{a priori} notion of time. Such a paradigm would also be more in line with canonical quantum gravity, for this latter is not formulated as a theory with Hamiltonian dynamics but as a set of states satisfying a (static) constraint equation \cite{DeWitt:1967yk, Wheeler:1968,Isham:1992ms}. In a (distantly) related spirit, we have derived constraints for whether or not a given process matrix may be interpreted as inducing a causal order between an emergent set of laboratories.

An exciting next step for this line of research would be to determine which additions to these constraints (and to the process matrix framework itself) would give rise to a full spacetime structure governed by gravitational equations. 
Others have explored the conditions for a given tensor network or a given operator algebra to be interpreted as a field theory on a background spacetime satisfying Einstein's equations~\cite{2018PhRvD..97h6003C,2026arXiv260610924M}. These results rely on the finding that localized agents perceive the first and second laws of thermodynamics thanks to gravity~\cite{1995PhRvL..75.1260J}. It seems reasonable to expect that a similar requirement for agents might be added in our program.

To move in the direction of empirical physics, one would need to supply the process-matrix formalism with dynamical structure. A further step would be to impose geometric constraints \cite{apadula2026frameperspectivesprocessmatrices}. A global unitary may manufacture a totally ordered setting, but it need not produce anything resembling a local spacetime. Geometric locality would therefore enter as an additional restriction on agents. Even if our results only bear on the total order between agents, which is not required for agents localized in Lorentzian manifolds, we speculate that spectral data should tell us whether and how one can define agents in a causally ordered process, while locality and geometry should help to distinguish agentic settings that are empirically realizable. 

\vspace{0.5cm}

\begin{acknowledgments}
{\it Code availability -- } A Julia package for quantum mereology is available at  \url{https://github.com/nicolasloizeau/QuantumMereology.jl}. We have also used and developed PauliStrings.jl: \url{https://paulistrings.org/} \cite{ps}.
\end{acknowledgments}
\vspace{0.5cm}

\begin{acknowledgments}
{\it Acknowledgments -- }
 V.K.\ gratefully acknowledges support from the \emph{Deutscher Akademischer Austauschdienst} (DAAD, German Academic Exchange Service). N.L. was supported by a research grant (42085) from Villum Fonden. N.L. thanks Berislav Buča for his support and for the freedom to pursue this project. A.G. was supported by l’Agence Nationale de la Recherche (ANR), project ANR-22-CE47-0012. O.F.\ was supported by a Fraunhofer-Schwarzschild-Fellowship at Universitätssternwarte München (LMU observatory) and by DFG's Excellence Cluster ORIGINS (EXC-2094/2 – 390783311).
\end{acknowledgments}

\bibliography{bib}

\appendix
\input{proofs}

\newpage
\let\oldaddcontentsline\addcontentsline
\renewcommand{\addcontentsline}[3]{}

\input{algorithms}

\end{document}

%% file: header.tex
\usepackage{amsmath, amssymb, mathtools}
\usepackage{amsthm}
\usepackage{bm}
\usepackage{bbold, mathrsfs}
\usepackage{graphicx, float}
\usepackage{comment}
\usepackage{braket}
\usepackage{algorithm, algpseudocode}
\usepackage{enumitem}
\usepackage{xcolor}
\usepackage{tcolorbox}
\tcbuselibrary{breakable, skins}
\usepackage{wrapfig}
\usepackage[colorlinks=true, allcolors=blue]{hyperref}
\theoremstyle{plain} 
\newtheorem{theorem}{Theorem} 
\newtheorem{proposition}{Proposition}

\theoremstyle{definition} 
 
\theoremstyle{remark}

\newcommand{\id}{\mathbb{1}}

\DeclareMathOperator{\tr}{tr}

\newcommand{\Ain}{A_I}
\newcommand{\Aout}{A_O}
\newcommand{\Bin}{B_I}
\newcommand{\Bout}{B_O}
\newcommand{\HAin}{\mathcal{H}_{A_I}}
\newcommand{\HAout}{\mathcal{H}_{A_O}}
\newcommand{\HBin}{\mathcal{H}_{B_I}}
\newcommand{\HBout}{\mathcal{H}_{B_O}}
\newcommand{\Htot}{\mathcal{H}}

\newcommand{\ocb}{W_\text{OCB}}
\newcommand{\AprecB}{A\prec B}
\newcommand{\BprecA}{B\prec A}

\newcommand{\process}[4]{#1_{A_I} #2_{A_O} #3_{B_I} #4_{B_O}}
\newcommand{\cP}{\mathcal{P}}
\newcommand{\cC}{\mathcal{C}}

\tcbset{
  algobox/.style={
    enhanced,
    breakable,
    colback=gray!5,
    colframe=black!70,
    fonttitle=\bfseries\small,
    title={#1},
    left=6pt, right=6pt, top=4pt, bottom=4pt,
    boxrule=0.5pt,
    arc=3pt,
  }
}

\newcommand{\ind}{\hspace{12pt}}

%% file: proofs.tex
\section{Proofs of the spectral results}
\label{app:spectral_orderability_proofs}
\subsection{Replacement maps and fixed-order factorization}

\label{app:fixed_order_factorization}

For a subsystem $X$\,, define
\begin{equation}
{}_{X}M\coloneqq\frac{\mathbb 1_X}{d_X}\otimes\tr_X M~, \qquad {}_{[1-X]}M\coloneqq M-{}_{X}M~.
\label{eq:replacement_map_app}
\end{equation}
When several systems appear as a subscript, the replacement acts on all of them. These maps give the standard linear constraints defining process matrices and fixed-order process classes \cite{Oreshkov2012,Oreshkov_2016,Branciard_2016_witness}.

For the total order $1\prec2\prec\cdots\prec n$\,, the last agent has no future laboratory. The corresponding fixed-order constraint is
\begin{equation}
{}_{[1-O_n]}W=0~.
\label{eq:last_party_constraint_app}
\end{equation}
Equivalently,
\begin{equation}
W={}_{O_n}W=\frac{\mathbb 1_{n,O}}{d_{n,O}}\otimes K~, \qquad K\coloneqq\tr_{O_n}W~.
\label{eq:last_factor_app}
\end{equation}
If $W\geq0$\,, then $K\geq0$\,. The remaining total order constraints are the recursive constraints on $K$ acting on
\begin{equation}
\mathcal K_{n-1}
\coloneqq
\left[
\bigotimes_{k=1}^{n-1}
\left(\mathcal H_{k,I}\otimes\mathcal H_{k,O}\right)
\right]
\otimes
\mathcal H_{n,I}~.
\label{eq:reduced_space_app}
\end{equation}
Explicitly, for $k=1,\dots,n-1$, let
\begin{equation}
F_k\coloneqq I_{k+1}O_{k+1}\cdots I_{n-1}O_{n-1}I_n
\end{equation}
denote the future systems of $O_k$ inside the reduced space, then 
\begin{equation}
{}_{[1-O_k]\,F_k}K=0~.
\label{eq:recursive_constraints_app}
\end{equation}
This proves Eq.~\eqref{eq:last_output_factorisation_main}. If $K$ has eigenvalue $\mu_a$ with multiplicity $r_a$\,, then $\mathbb 1_{n,O}\otimes K/d_{n,O}$ has eigenvalue $\mu_a/d_{n,O}$ with multiplicity $d_{n,O}r_a$\,, giving Proposition~\ref{prop:spectral_obstruction}.

\subsection{Sufficiency of the spectral criterion}
\label{app:sufficiency_spectral_condition}

We prove the sufficiency direction of Theorem~\ref{thm:exact_characterization}. The idea is easier to see before writing the general case, so we first give the bipartite qubit construction. The general proof is then the same construction repeated along the ordered chain.
\subsubsection*{Warm-up: two qubit laboratories}

Consider two qubit laboratories in the order
\begin{equation}
A\prec B~.
\end{equation}
The process Hilbert space is
\begin{equation}
\mathcal H=
\mathcal H_{A_I}\otimes\mathcal H_{A_O}\otimes
\mathcal H_{B_I}\otimes\mathcal H_{B_O}~,
\end{equation}
with all four factors two-dimensional. The output normalization is
\begin{equation}
D_O=d_{A_O}d_{B_O}=4~.
\end{equation}

A process in the order $A\prec B$ has Bob as the last agent. Hence Bob's output is an open final wire:
\begin{equation}
{}_{[1-B_O]}X=0~.
\end{equation}
Equivalently,
\begin{equation}
X=\frac{\mathbb 1_{B_O}}{2}\otimes K~,
\label{eq:bipartite_qubit_order_ansatz_app}
\end{equation}
where $K$ acts on $A_IA_OB_I$. The remaining reduced fixed-order constraint is
\begin{equation}
{}_{[1-A_O]}\,{}_{B_I}K=0~.
\label{eq:bipartite_reduced_constraint_app}
\end{equation}
It says that, after the future input $B_I$ is replaced, Alice's output $A_O$ is maximally mixed.

Now let $W\geq0$ have $\tr W=4$ and suppose its spectrum is pair-degenerate:
\begin{equation}
\operatorname{spec}(W)=
(\lambda_1,\lambda_1,\lambda_2,\lambda_2,\ldots,\lambda_8,\lambda_8)~.
\label{eq:bipartite_paired_spectrum_app}
\end{equation}
If $X$ has the form in Eq.~\eqref{eq:bipartite_qubit_order_ansatz_app}, then every eigenvalue of $K$ is divided by $2$ and repeated twice in $X$. Thus, to match the spectrum of $W$, we should choose $K$ to have eigenvalues
\begin{equation}
2\lambda_1,\;2\lambda_2,\;\ldots,\;2\lambda_8~.
\label{eq:bipartite_reduced_eigenvalues_app}
\end{equation}
The spectrum fixes these eigenvalues, but it does not fix the eigenvectors. We now choose the eigenvectors so that the reduced fixed-order constraint holds automatically.

Let
\begin{equation}
\begin{aligned}
\ket{\Phi^\pm}
&=\frac{1}{\sqrt2}\bigl(\ket{00}\pm\ket{11}\bigr)~,\\
\ket{\Psi^\pm}
&=\frac{1}{\sqrt2}\bigl(\ket{01}\pm\ket{10}\bigr)~.
\end{aligned}
\label{eq:bell_basis_app}
\end{equation}
be the Bell basis on $A_OB_I$. Each Bell state $\ket{\beta}$ satisfies
\begin{equation}
\tr_{B_I}\ket{\beta}\bra{\beta}=\frac{\mathbb 1_{A_O}}{2}~.
\label{eq:bell_marginal_app}
\end{equation}
Choose the computational basis $\{\ket{0},\ket{1}\}$ of $A_I$ and define
\begin{equation}
\begin{aligned}
\ket{e_{a,\beta}}
&=
\ket{a}_{A_I}\otimes\ket{\beta}_{A_OB_I}~,\\
a&=0,1~,
\qquad
\beta\in\{\Phi^+,\Phi^-,\Psi^+,\Psi^-\}~.
\end{aligned}
\label{eq:bipartite_bridge_basis_app}
\end{equation}
These eight vectors form an orthonormal basis of $A_IA_OB_I$. Assign the eigenvalues in Eq.~\eqref{eq:bipartite_reduced_eigenvalues_app} to this basis and set
\begin{equation}
\widetilde K=\sum_{a,\beta}\kappa_{a,\beta}\ket{e_{a,\beta}}\bra{e_{a,\beta}}~,
\label{eq:bipartite_K_tilde_app}
\end{equation}
where $\{\kappa_{a,\beta}\}=\{2\lambda_1,\dots,2\lambda_8\}$, counted with multiplicity.

Let $\mathbb P_{a,\beta}=\ket{e_{a,\beta}}\bra{e_{a,\beta}}$. Using Eq.~\eqref{eq:bell_marginal_app},
\begin{equation}
{}_{B_I}\mathbb P_{a,\beta}
=
\ket{a}\bra{a}_{A_I}
\otimes
\frac{\mathbb 1_{A_O}}{2}
\otimes
\frac{\mathbb 1_{B_I}}{2}~.
\end{equation}
The right-hand side is already maximally mixed on $A_O$, so
\begin{equation}
{}_{[1-A_O]}\,{}_{B_I}\mathbb P_{a,\beta}=0~.
\end{equation}
By linearity,
\begin{equation}
{}_{[1-A_O]}\,{}_{B_I}\widetilde K=0~.
\end{equation}
Therefore
\begin{equation}
X=\frac{\mathbb 1_{B_O}}{2}\otimes\widetilde K
\label{eq:bipartite_X_constructed_app}
\end{equation}
is a valid process in $\mathscr W_{A\prec B}$. Indeed, $X\geq0$, $\tr X=\tr\widetilde K=\tr W=4$, ${}_{[1-B_O]}X=0$, and the remaining reduced constraint was just checked. Finally, the factor $\mathbb 1_{B_O}/2$ turns the eigenvalues $2\lambda_j$ of $\widetilde K$ into the paired eigenvalues $\lambda_j,\lambda_j$ of $X$. Hence
\begin{equation}
\operatorname{spec}(X)=\operatorname{spec}(W)~.
\end{equation}
Thus $W=UXU^\dagger$ for some global unitary $U$, and $W\in\mathcal O_{A\prec B}$.

This is the whole mechanism. The spectrum tells us which eigenvalues the reduced operator should have. The Bell basis supplies eigenvectors whose projectors satisfy the fixed total-order constraints. The final identity factor supplies the required spectral degeneracy.

\subsubsection*{General construction}

We now repeat the same construction for the total order
\begin{equation}
1\prec2\prec\cdots\prec n~.
\end{equation}
A process in this order has the factorization
\begin{equation}
X=\frac{\mathbb 1_{n,O}}{d_{n,O}}\otimes K~,
\label{eq:general_order_ansatz_app}
\end{equation}
where $K$ acts on the reduced Hilbert space
\begin{equation}
\mathcal K_{n-1}
\coloneqq
\left[
\bigotimes_{k=1}^{n-1}
\left(\mathcal H_{k,I}\otimes\mathcal H_{k,O}\right)
\right]
\otimes
\mathcal H_{n,I}~.
\label{eq:general_reduced_space_app}
\end{equation}
The remaining fixed-order constraints are
\begin{equation}
{}_{[1-O_k]}\,{}_{F_k}K=0~,
\qquad
k=1,\dots,n-1~.
\label{eq:general_reduced_constraints_app}
\end{equation}
The construction uses the chain-divisibility assumption
\begin{equation}
d_{k,O}\mid d_{k+1,I}~, \qquad k=1,\dots,n-1~.
\label{eq:chain_divisibility_app}
\end{equation}
Write $d_{k+1,I}=q_kd_{k,O}$. Then each future input can be decomposed into blocks of the same dimension as the previous output:
\begin{equation}
\mathcal H_{k+1,I}
=
\bigoplus_{r=1}^{q_k}\mathcal K_r^{(k)}~,
\qquad
\dim\mathcal K_r^{(k)}=d_{k,O}~.
\label{eq:input_decomposition_app}
\end{equation}
Let
\begin{equation}
E_r^{(k)}:\mathcal H_{k,O}\longrightarrow
\mathcal K_r^{(k)}\subset\mathcal H_{k+1,I}
\end{equation}
be the isometry into the $r$-th block. This is where Eq.~\eqref{eq:chain_divisibility_app} is used: it allows $O_k$ to be paired with a block of $I_{k+1}$ in a Bell-like basis.

Choose an orthonormal basis $\{\ket{j}\}_{j=0}^{d_{k,O}-1}$ of $\mathcal H_{k,O}$, and write
\begin{equation}
\ket{r,j}:=E_r^{(k)}\ket{j}~.
\end{equation}
Let $\omega_k=e^{2\pi i/d_{k,O}}$ and define the Weyl operators
\begin{equation}
Z_k\ket{j}=\omega_k^j\ket{j}~, 
\quad
X_k\ket{j}=\ket{j+1\!\!\!\pmod {d_{k,O}}}~,
\end{equation}
with
\begin{equation}
U_{a,b}^{(k)}\coloneqq X_k^bZ_k^a~,
\qquad
a,b=0,\dots,d_{k,O}-1~.
\label{eq:weyl_operator_basis_app}
\end{equation}
These unitaries obey
\begin{equation}
\tr\!\left[
\left(U_{a,b}^{(k)}\right)^\dagger
U_{a',b'}^{(k)}
\right]
=
d_{k,O}\,\delta_{a,a'}\delta_{b,b'}~,
\label{eq:weyl_orthogonality_app}
\end{equation}
so they form a Hilbert--Schmidt orthogonal operator basis of $\mathcal B(\mathcal H_{k,O})$ \cite{Knill1996ErrorBases,Watrous2018}.

For a map $V:\mathcal H_{k,O}\to\mathcal H_{k+1,I}$, use the vectorization convention
\begin{equation}
\ket{V}\!\rangle\coloneqq
\sum_{j=0}^{d_{k,O}-1}\ket{j}_{k,O}\otimes V\ket{j}_{k+1,I}~.
\label{eq:vectorization_convention_app}
\end{equation}
Define the bridge states
\begin{equation}
\ket{\eta_{r,a,b}^{(k)}}\coloneqq
\frac{1}{\sqrt{d_{k,O}}}
\ket{E_r^{(k)}U_{a,b}^{(k)}}\!\rangle~.
\label{eq:bridge_state_vectorized_app}
\end{equation}
Equivalently,
\begin{equation}
\begin{aligned}
\ket{\eta_{r,a,b}^{(k)}}
&=
\frac{1}{\sqrt{d_{k,O}}}
\sum_{j=0}^{d_{k,O}-1}
\omega_k^{aj}\,
\ket{j}_{k,O} \\
&\hspace{2.5cm}\otimes
\ket{r,j+b}_{k+1,I}~,
\end{aligned}
\label{eq:bridge_state_explicit_app}
\end{equation}
where $j+b$ is taken modulo $d_{k,O}$. For $d_{k,O}=2$ and $q_k=1$, these are exactly the four Bell states.

The bridge states have two properties. First, they form a full orthonormal basis of $\mathcal H_{k,O}\otimes\mathcal H_{k+1,I}$:
\begin{equation}
\braket{\eta_{r,a,b}^{(k)}\mid{\eta_{r',a',b'}^{(k)}}}
=
\delta_{r,r'}\delta_{a,a'}\delta_{b,b'}~.
\label{eq:bridge_orthogonality_app}
\end{equation}
Indeed, different values of $r$ land in orthogonal blocks, while for fixed $r$ this is just Eq.~\eqref{eq:weyl_orthogonality_app}. The number of states is
\begin{equation}
q_k d_{k,O}^2=d_{k,O}d_{k+1,I}
=
\dim(\mathcal H_{k,O}\otimes\mathcal H_{k+1,I})~,
\end{equation}
so the orthonormal set is complete.

Second, every bridge state is maximally mixed on the earlier output:
\begin{equation}
\tr_{k+1,I}
\ket{\eta_{r,a,b}^{(k)}}\bra{\eta_{r,a,b}^{(k)}}
=
\frac{\mathbb 1_{k,O}}{d_{k,O}}~.
\label{eq:bridge_marginal_app}
\end{equation}
This follows directly from Eq.~\eqref{eq:bridge_state_explicit_app}: tracing over $I_{k+1}$ removes the off-diagonal terms and leaves the uniform diagonal operator on $O_k$.

Now rearrange the reduced Hilbert space as
\begin{equation}
\mathcal K_{n-1}
\cong
\mathcal H_{1,I}
\otimes
\bigotimes_{k=1}^{n-1}
\left(
\mathcal H_{k,O}\otimes\mathcal H_{k+1,I}
\right)~.
\label{eq:rearranged_reduced_space_app}
\end{equation}
Choose any orthonormal basis $\{\ket{\alpha}\}$ of $\mathcal H_{1,I}$ and tensor it with the bridge bases for all pairs $O_kI_{k+1}$. This gives an orthonormal basis $\{\ket{e_J}\}$ of $\mathcal K_{n-1}$, with typical element
\begin{equation}
\ket{e_J}
=
\ket{\alpha}_{1,I}
\otimes
\bigotimes_{k=1}^{n-1}
\ket{\eta^{(k)}}_{k,O\,\,k+1,I}~.
\label{eq:global_bridge_basis_vector_app}
\end{equation}
Here $J$ is a collective label for $\alpha$ and for all bridge labels.
Let $\mathbb P_J=\ket{e_J}\bra{e_J}$. Fix $k<n$. In the constraint Eq.~\eqref{eq:general_reduced_constraints_app}, the future replacement traces over $I_{k+1}$, among other systems. The factor of $\mathbb P_J$ on $O_kI_{k+1}$ is a bridge projector, and Eq.~\eqref{eq:bridge_marginal_app} leaves $\mathbb 1_{k,O}/d_{k,O}$. Thus $O_k$ is already maximally mixed after the future has been replaced:
\begin{equation}
{}_{[1-O_k]}\,
{}_{I_{k+1}O_{k+1}\cdots I_{n-1}O_{n-1}I_n}
\mathbb P_J=0~.
\end{equation}
By linearity, every operator diagonal in the basis $\{\ket{e_J}\}$ satisfies all reduced fixed-order constraints.

We now choose the diagonal entries from the spectrum of $W$. Let $d_f=d_{n,O}$. Since every distinct eigenvalue multiplicity of $W$ is divisible by $d_f$, write
\begin{equation}
\operatorname{spec}(W)
=
(\underbrace{\lambda_1,\dots,\lambda_1}_{d_f},
 \ldots, \underbrace{\lambda_M,\dots,\lambda_M}_{d_f})~,
\label{eq:grouped_spectrum_app}
\end{equation}
where repeated values among the $\lambda_j$'s are allowed. Define
\begin{equation}
\kappa_j=d_f\lambda_j~,
\qquad
j=1,\dots,M~,
\end{equation}
and set
\begin{equation}
\widetilde K=\sum_{J=1}^{M}\kappa_J\ket{e_J}\bra{e_J}~.
\label{eq:K_tilde_general_app}
\end{equation}
Then $\widetilde K\geq0$, it has the required reduced spectrum, and it satisfies all reduced fixed-order constraints. Moreover,
\begin{equation}
\tr\widetilde K
=
\sum_{j=1}^{M}d_f\lambda_j
=
\tr W
=
D_O~.
\end{equation}

Finally define
\begin{equation}
X=\frac{\mathbb 1_{n,O}}{d_f}\otimes\widetilde K~.
\label{eq:constructed_ordered_process_app}
\end{equation}
Then $X\geq0$, $\tr X=D_O$, the final output is open, and the remaining fixed total-order constraints hold by construction. Hence
\begin{equation}
X\in\mathscr W_{1\prec\cdots\prec n}~.
\end{equation}
The spectrum also matches: each eigenvalue $\kappa_j=d_f\lambda_j$ of $\widetilde K$ becomes $\lambda_j$ and is repeated $d_f$ times in $X$. Therefore
\begin{equation}
\operatorname{spec}(X)=\operatorname{spec}(W)~.
\end{equation}
Since finite-dimensional Hermitian matrices with the same spectrum are unitarily equivalent, there exists $U\in\mathsf U(\mathcal H)$ such that
\begin{equation}
W=UXU^\dagger~.
\end{equation}
Thus $W\in\mathcal O_{1\prec\cdots\prec n}$, completing the proof.

\subsection{Agent redefinition by splitting and merging}
\label{app:agent_redefinition}

Let the global Hilbert space be built from $N$ elementary input-output pairs labelled by $1,\ldots,N$. A decomposition into $m$ agents is specified by a partition
\begin{equation}
\{1,\ldots,N\} = G_1\sqcup G_2\sqcup\cdots\sqcup G_m
\label{eq:agent_partition_app}
\end{equation}
together with a total order
\begin{equation}
\widetilde A_1 \prec \widetilde A_2 \prec\cdots\prec \widetilde A_m~.
\end{equation}
The agent $\widetilde A_a$ is defined by the set $G_a$, with
\begin{equation}
\begin{aligned}
\mathcal H_{\widetilde A_a,I} &\coloneqq \bigotimes_{k\in G_a}\mathcal H_{k,I}~,\\
\mathcal H_{\widetilde A_a,O} &\coloneqq \bigotimes_{k\in G_a}\mathcal H_{k,O}~.
\end{aligned}
\label{eq:grouped_agent_spaces_app}
\end{equation}
Its input and output dimensions are
\begin{equation}
\begin{aligned}
d_{\widetilde A_a,I} &= \prod_{k\in G_a}d_{k,I}~,\\
d_{\widetilde A_a,O} &= \prod_{k\in G_a}d_{k,O}~.
\end{aligned}
\label{eq:grouped_agent_dimensions_app}
\end{equation}

Splitting an agent $\widetilde A_a$ means choosing a further partition
\begin{equation}
G_a = H_{a,1}\sqcup\cdots\sqcup H_{a,s}
\end{equation}
and replacing $\widetilde A_a$ by $s$ agents defined by the sets
$H_{a,1},\ldots,H_{a,s}$, with some chosen order among them. Merging consecutive agents
$\widetilde A_a,\ldots,\widetilde A_b$ means replacing them by one agent defined by
\begin{equation}
H = G_a\sqcup G_{a+1}\sqcup\cdots\sqcup G_b~.
\end{equation}
Both operations leave the total Hilbert space unchanged.

Let $\widetilde L=\widetilde A_m$ be the last agent. Its output dimension is
\begin{equation}
d_{\widetilde L,O} = \dim\mathcal H_{\widetilde L,O} = \prod_{k\in G_m}d_{k,O}~.
\label{eq:last_agent_output_dimension_app}
\end{equation}
Applying Theorem~\ref{thm:exact_characterization} to this decomposition gives
\begin{equation}
W\in \mathcal O_{\widetilde A_1\prec\cdots\prec\widetilde A_m} \quad\Longleftrightarrow\quad d_{\widetilde L,O}\mid g(W)~,
\label{eq:agent_redefinition_condition_app}
\end{equation}
provided that
\begin{equation}
d_{\widetilde A_a,O} \mid d_{\widetilde A_{a+1},I}~, \qquad a=1,\ldots,m-1~.
\label{eq:grouped_chain_divisibility_app}
\end{equation}
Without this dimension assumption, divisibility by $d_{\widetilde L,O}$ remains necessary, but Theorem~\ref{thm:exact_characterization} does not guarantee that it is sufficient.

Only changes involving the last agent alter the divisor in
Eq.~\eqref{eq:agent_redefinition_condition_app} directly. If the last agent is split as
\begin{equation}
G_m = H_1\sqcup\cdots\sqcup H_s
\end{equation}
and the agents defined by $H_1,\ldots,H_s$ are placed last in this order, then the new final-output divisor is
\begin{equation}
d_{H_s,O} = \prod_{k\in H_s}d_{k,O}~.
\end{equation}
This may be smaller than $d_{\widetilde L,O}$. Conversely, merging several final agents replaces their output dimensions by their product and therefore strengthens the required divisibility. In either case, the chain-divisibility condition must be checked again for the new ordering.

In the equal-dimensional case,
\begin{equation}
d_{k,I}=d_{k,O}=d~,
\end{equation}
an agent defined by $G_a$ has
\begin{equation}
d_{\widetilde A_a,I} = d_{\widetilde A_a,O} = d^{|G_a|}~.
\end{equation}
The chain-divisibility condition then reduces to
\begin{equation}
|G_a| \leq |G_{a+1}|~, \qquad a=1,\ldots,m-1~.
\label{eq:group_size_condition_app}
\end{equation}
If the last agent contains
\begin{equation}
b=|G_m|
\end{equation}
elementary input-output pairs, then
\begin{equation}
d_{\widetilde L,O}=d^b~.
\end{equation}
Using Eq.~\eqref{eq:divisibility_depth}, the spectral condition becomes
\begin{equation}
W\in \mathcal O_{\widetilde A_1\prec\cdots\prec\widetilde A_m} \quad\Longleftrightarrow\quad b\leq\ell_d(W)~,
\label{eq:last_agent_depth_rule_app}
\end{equation}
provided that Eq.~\eqref{eq:group_size_condition_app} also holds. Thus $\ell_d(W)$ constrains the number of elementary output factors assigned to the last agent, while the sizes and ordering of the remaining agents are constrained separately by chain divisibility.

\subsection{Distance to the fixed-order orbit}
\label{app:distance_formula}
Let $\rho$ have eigenvalues
\begin{equation}
\nu_1\ge\cdots\ge\nu_D\ge0~, \qquad \sum_{i=1}^D\nu_i=1~.
\end{equation}
For the total order $\pi$, write $N_\pi=\frac{D}{d_{\pi(n),O}}$\,. By Theorem~\ref{thm:exact_characterization}, a normalized operator lies in $\widehat{\mathcal O}_\pi$ only if its eigenvalues are constant in consecutive blocks of length $d_{\pi(n),O}$. Thus a spectrum in the fixed-order orbit has the form
\begin{equation}
\underbrace{x_1,\ldots,x_1}_{d_{\pi(n),O}\ \mathrm{times}}, \underbrace{x_2,\ldots,x_2}_{d_{\pi(n),O}\ \mathrm{times}}~,
\ldots, \underbrace{x_{N_\pi},\ldots,x_{N_\pi}}_{d_{\pi(n),O}\ \mathrm{times}}~,
\end{equation}
with $x_a\ge x_{a+1}$ and $d_{\pi(n),O}\sum_{a=1}^{N_\pi}x_a=1$\,.

The Hilbert--Schmidt distance between two Hermitian matrices is minimized, for fixed spectra, by diagonalizing them in the same eigenbasis and matching the eigenvalues in decreasing order. This is the Hoffman--Wielandt, or spectral rearrangement, principle for the Frobenius norm \cite{HoffmanWielandt1953,Bhatia1997}. The problem is therefore to choose the numbers $x_a$ so as to minimize
\begin{equation}
\sum_{a=1}^{N_\pi}\sum_{j\in B_a}(\nu_j-x_a)^2~,
\end{equation}
where
\begin{equation}
B_a=\{(a-1)d_{\pi(n),O}+1,\ldots,ad_{\pi(n),O}\}~.
\end{equation}

This minimization is independent from block to block. For a fixed block $B_a$, the best constant approximation to the eigenvalues in that block is their average. Indeed,
\begin{equation}
\frac{\rm d}{\rm d\, x_a} \sum_{j\in B_a}(\nu_j-x_a)^2 = -2\sum_{j\in B_a}(\nu_j-x_a)~,
\end{equation}
which vanishes at
\begin{equation}
x_a=\bar\nu_a \coloneqq \frac{1}{d_{\pi(n),O}}\sum_{j\in B_a}\nu_j~.
\end{equation}
Thus the closest block-degenerate spectrum is obtained by replacing each block by its average.

Since the original eigenvalues are nonincreasing, the block averages are also nonincreasing:
\begin{equation}
\bar\nu_1\ge\bar\nu_2\ge\cdots\ge\bar\nu_{N_\pi}~.
\end{equation}
The ordering constraint is therefore automatically satisfied. Moreover, the normalization is preserved, because
\begin{equation}
d_{\pi(n),O}\sum_{a=1}^{N_\pi}\bar\nu_a = \sum_{i=1}^D\nu_i = 1~.
\end{equation}
Hence the squared distance to the fixed-order orbit is
\begin{equation}
\widehat\delta^{(2)}_\pi(\rho)^2 = \sum_{a=1}^{N_\pi} \sum_{j\in B_a} (\nu_j-\bar\nu_a)^2~.
\label{eq:distance_block_variance_app}
\end{equation}
Expanding the square in each block gives
\begin{equation}
\begin{aligned}
\sum_{j\in B_a}(\nu_j-\bar\nu_a)^2 &= \sum_{j\in B_a}\nu_j^2 - d_{\pi(n),O}\bar\nu_a^2~.
\end{aligned}
\end{equation}
Summing over all blocks yields
\begin{equation}
\widehat\delta^{(2)}_\pi(\rho)^2 = \tr(\rho^2) - d_{\pi(n),O}\sum_{a=1}^{N_\pi}\bar\nu_a^2~,
\end{equation}
which is Eq.~\eqref{eq:distance_formula}. The purity bound Eq.~\eqref{eq:purity_bound} follows immediately by dropping the nonnegative term $d_{\pi(n),O}\sum_a\bar\nu_a^2$\,.

\subsection{Obstruction for same-time unitaries}
\label{app:same_time_obstruction}

We prove Proposition~\ref{prop:sametime_obstruction}. A same-time unitary may redefine the agents inside the input slice and inside the output slice, but it cannot mix inputs with outputs. The obstruction below comes from this restriction.

Let
\begin{equation}
\mathcal H_I\coloneqq\mathcal H_{A_I}\otimes\mathcal H_{B_I}~,
\qquad
\mathcal H_O\coloneqq\mathcal H_{A_O}\otimes\mathcal H_{B_O}~,
\end{equation}
and let
\begin{equation}
U_{\rm st}=U_I\otimes U_O~,
\quad
U_I\in\mathsf U(\mathcal H_I)~,
\quad
U_O\in\mathsf U(\mathcal H_O)~.
\end{equation}
Thus $U_I$ acts only on the input slice and $U_O$ acts only on the output slice.

We first expand $W$ across the input-output cut. Let $d_I=\dim\mathcal H_I$, and choose a Hilbert--Schmidt orthonormal basis $A_1,\dots,A_{d_I^2-1}$ of traceless Hermitian operators on $\mathcal H_I$. Together with $\mathbb 1_I/\sqrt{d_I}$, this is an orthonormal basis of Hermitian input operators. Hence every Hermitian operator $W$ on $\mathcal H_I\otimes\mathcal H_O$ can be written as
\begin{equation}
W= \frac{\mathbb 1_I}{d_I}\otimes C_0 + \sum_{k=1}^{d_I^2-1}A_k\otimes B_k~,
\label{eq:io_decomposition_same_time_app}
\end{equation}
where
\begin{equation}
C_0=\tr_I W~, \qquad B_k=\tr_I\!\left[(A_k\otimes\mathbb 1_O)W\right]~.
\end{equation}
The first term is the input-independent part of $W$. The remaining terms contain all nontrivial input-output dependence. We define the associated output operator system by
\begin{equation}
\begin{aligned}
\mathcal S_{\rm out}(W)
&\coloneqq
\operatorname{span}\{B_k:1\le k\le d_I^2-1\}\\
&\subset \mathcal B(\mathcal H_O)~.
\end{aligned}
\label{eq:S_out_same_time_app}
\end{equation}

Suppose that a same-time unitary reveals the order $\widetilde A\prec\widetilde B$. Then $\widetilde B_O$ is the final output in the transformed description, so
\begin{equation}
{}_{[1-\widetilde B_O]}\, U_{\rm st}WU_{\rm st}^\dagger= 0~.
\label{eq:final_output_condition_same_time_app}
\end{equation}
Define
\begin{equation}
\begin{aligned}
A'_k&\coloneqq U_I A_k U_I^\dagger~,\\
B'_k&\coloneqq U_O B_k U_O^\dagger~,\\
C'_0&\coloneqq U_O C_0 U_O^\dagger~.
\end{aligned}
\label{eq:transformed_coefficients_same_time_app}
\end{equation}
Then
\begin{equation}
U_{\rm st}WU_{\rm st}^\dagger =\frac{\mathbb 1_I}{d_I}\otimes C'_0+\sum_{k=1}^{d_I^2-1}A'_k\otimes B'_k~.
\end{equation}
The operators $A'_k$ are still Hilbert--Schmidt orthonormal and traceless. Applying Eq.~\eqref{eq:final_output_condition_same_time_app} gives
\begin{equation}
0 =\frac{\mathbb 1_I}{d_I}\otimes {}_{[1-\widetilde B_O]}C'_0+ \sum_{k=1}^{d_I^2-1}A'_k\otimes {}_{[1-\widetilde B_O]}B'_k~.
\label{eq:expanded_final_output_condition_app}
\end{equation}
Now take the Hilbert--Schmidt inner product with $A'_j$ on the input side. The first term drops out because $\tr A'_j=0$, and orthonormality isolates the $j$-th coefficient. Therefore
\begin{equation}
{}_{[1-\widetilde B_O]}B'_j=0~, \qquad j=1,\dots,d_I^2-1~.
\label{eq:output_coefficients_final_mixed_app}
\end{equation}
Thus every transformed output coefficient is already maximally mixed on $\widetilde B_O$. Equivalently,
\begin{equation}
U_O\,\mathcal S_{\rm out}(W)\,U_O^\dagger \subset \mathcal B(\mathcal H_{\widetilde A_O})\otimes\mathbb 1_{\widetilde B_O}~.
\label{eq:output_inclusion_A_before_B_app}
\end{equation}
So if a same-time unitary reveals the order $\widetilde A\prec\widetilde B$, all nontrivial output dependence must fit inside the output algebra of the earlier party. Repeating the same argument for the opposite order $\widetilde B\prec\widetilde A$ gives
\begin{equation}
U_O\,\mathcal S_{\rm out}(W)\,U_O^\dagger \subset \mathbb 1_{\widetilde A_O}\otimes\mathcal B(\mathcal H_{\widetilde B_O})~.
\label{eq:output_inclusion_B_before_A_app}
\end{equation}

We now use a simple dimension bound. Any commuting family of Hermitian matrices is simultaneously diagonalizable. Hence a commuting traceless Hermitian subspace of $d\times d$ matrices has dimension at most $d-1$, because traceless diagonal matrices form a real vector space of dimension $d-1$.

Suppose $\mathcal S_{\rm out}(W)$ contains a commuting traceless Hermitian subspace of dimension $q$. Unitary conjugation preserves dimension, Hermiticity, tracelessness, and commutativity. If the inclusion \eqref{eq:output_inclusion_A_before_B_app} held, this subspace would have to fit inside $\mathcal B(\mathcal H_{\widetilde A_O})\otimes\mathbb 1_{\widetilde B_O}$, whose largest commuting traceless Hermitian subspace has dimension $d_{A_O}-1$. Similarly, if the inclusion \eqref{eq:output_inclusion_B_before_A_app} held, it would have to fit inside $\mathbb 1_{\widetilde A_O}\otimes\mathcal B(\mathcal H_{\widetilde B_O})$, whose corresponding dimension is $d_{B_O}-1$. The same-time unitary changes the decomposition but not the output dimensions, so these are the same $d_{A_O}$ and $d_{B_O}$ as before.

Therefore, if
\begin{equation}
q>\max(d_{A_O}-1,d_{B_O}-1)~,
\end{equation}
then neither inclusion is possible. No same-time unitary can transform $W$ into a definite-order process. This proves Proposition~\ref{prop:sametime_obstruction}.

%% file: algorithms.tex
\section{Numerical Methods}\label{app:numerical}

Here we describe numerical methods we use to solve the following problem : given a non causally ordered process matrix $W$, find a unitary $U$ such that $UWU^\dagger$ is ordered.
These methods also apply to more general mereological problems \cite{Loizeau2023, Loizeau2024}.

Let $H$ be an $n$-qubit Hermitian operator. Given a target set of Pauli strings
$\{P_i\}$ and a generating set $\{G_k\}$, all three methods seek a unitary $U$,
constrained to $\mathrm{span}\{G_k\}$, such that $UHU^\dagger$ is supported on
$\mathrm{span}\{P_i\}$. They differ in how $U$ is parameterized and optimized.

Define the inner product $\langle A,B\rangle = 2^{-n}\mathrm{tr}(A^\dagger B)$,
the projection $P_S(O)=\sum_i\langle P_i,O\rangle P_i$, and the spectral cost
\begin{equation}
    C(\mathbf{h}) = \|\lambda(H) - \lambda({\textstyle\sum_i} h_i P_i)\|^2,
\end{equation}
where $\lambda(\cdot)$ denotes the sorted eigenvalue vector.

\paragraph{Hamiltonian flow.}
We maximize the support overlap
$\mathcal{F}(U)=\|P_S(UHU^\dagger)\|^2$
by gradient ascent on the unitary manifold.
The gradient with respect to the generator coefficients $\theta_k$,
obtained from the first-order perturbation $H\to H-i\epsilon[G_k,H]$, is
\begin{equation}
    g_k = 2\,\mathrm{Re}\!\left(i\sum_i
          \langle P_i,[G_k,H]\rangle\,\langle P_i,H\rangle^*\right).
\end{equation}

\begin{figure}[H]
\centering
\includegraphics[width=0.4\textwidth]{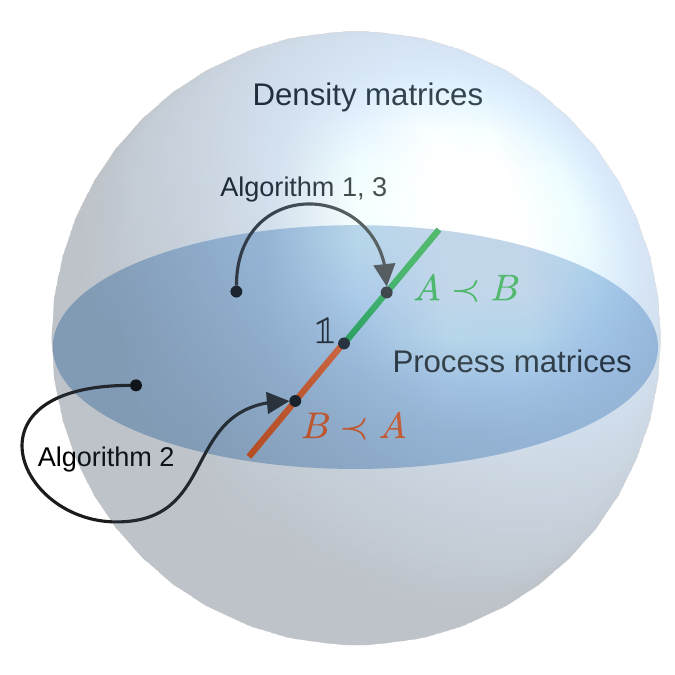}
\caption{Representation of the optimization routes the different methods described in this section take in the space of process matrices. The space of process matrices satisfying $\AprecB$ is a convex subset of the space of the process matrices, itself a convex subset of the space of the density matrices. Algorithms 1 and 3 exit the space of the valid process matrices during the optimization, but stay in the space of density matrices because they conserve the spectrum. Algorithm 2 does not conserve the spectrum during the procedure, and therefore it can exit the space of density matrices. }
\label{fig:sphere}
\end{figure}

\begin{tcolorbox}[algobox={Algorithm 1: Hamiltonian flow}]
\small
\textbf{Input:} $H$, target strings $\{P_i\}$, generators $\{G_k\}$, step $\alpha$\\
\textbf{Output:} $U$ such that $UHU^\dagger \approx P_S(UHU^\dagger)$
\tcblower
$U \gets \mathbf{1}$\\
\textbf{for} each iteration \textbf{do}\\
\ind compute $g_k = 2\,\mathrm{Re}(i\sum_i \langle P_i,[G_k,H]\rangle\langle P_i,H\rangle^*)$\\
\ind rescale $g_k \gets g_k / (\|[G_k,H]\|+\varepsilon)$\\
\ind normalize $\hat{g} \gets g/\|g\|$; add small noise\\
\ind find $\alpha$ by line search maximizing $\mathcal{F}$\\
\ind $U_i \gets \exp(-i\alpha\sum_k \hat{g}_k G_k)$\\
\ind $H \gets U_i H U_i^\dagger$,\quad $U \gets U_i U$\\
\textbf{end}\\
\textbf{return} $U$
\end{tcolorbox}

\paragraph{Gradient descent on spectral cost.}
Instead of acting on $U$ directly, we find the coefficients $\mathbf{h}=(h_i)$ of the
operator $H'=\sum_i h_i P_i$ whose spectrum best matches that of $H$, then
read off $U$ from the change of eigenbasis.

\begin{tcolorbox}[algobox={Algorithm 2: Gradient descent}]
\small
\textbf{Input:} $H$, target strings $\{P_i\}$\\
\textbf{Output:} $U$ such that $UHU^\dagger\approx\sum_i h_i P_i$
\tcblower
$(E, V) \gets \mathrm{Eig}(H)$\\
minimize $C(\mathbf{h})=\|E - \lambda(\sum_i h_i P_i)\|^2$ over $\mathbf{h}$\\
$(E', W) \gets \mathrm{Eig}(\sum_i h_i P_i)$\\
$U \gets V^\dagger W$\\
\textbf{return} $U$
\end{tcolorbox}

\paragraph{Projected gradient descent on spectral cost.}
The same spectral cost is minimized, but the unitary is constrained to
$\mathrm{span}\{G_k\}$ via a projection step at each iteration.

\begin{tcolorbox}[algobox={Algorithm 3: Projected gradient descent}]
\small
\textbf{Input:} $H$, target strings $\{P_i\}$, generators $\{G_k\}$, mixing $\alpha$\\
\textbf{Output:} $U$ such that $UHU^\dagger \approx P_S(UHU^\dagger)$, $U\in\langle G_k\rangle$
\tcblower
$U \gets \mathbf{1}$\\
\textbf{for} each iteration \textbf{do}\\
\ind $H' \gets \alpha\, P_S(H) + (1-\alpha)\,H$\\
\ind $(E_1,V_1)\gets\mathrm{Eig}(H)$;\quad $(E_2,V_2)\gets\mathrm{Eig}(H')$\\
\ind $U_i \gets V_2 V_1^\dagger$\\
\ind $G \gets \log(U_i)$;\quad $G \gets P_{\{G_k\}}(G)$;\quad $U_i \gets e^{G}$\\
\ind $H \gets U_i H U_i^\dagger$,\quad $U \gets U_i U$\\
\textbf{end}\\
\textbf{return} $U$
\end{tcolorbox}